\documentclass{emulateapj}

\shorttitle{Obscured QSOs in the COSMOS field}
\shortauthors{Fiore et al.}

\def\ls{{_<\atop^{\sim}}}
\def\gs{{_>\atop^{\sim}}}
\def\cgs{ ${\rm ergs~cm}^{-2}~{\rm s}^{-1}$ }
\def\cm2{cm$^{-2}$}

\begin{document}

\title{Chasing highly obscured QSOs in the COSMOS field.}
\author{F. Fiore\altaffilmark{1}, 
S. Puccetti\altaffilmark{2},
M. Brusa\altaffilmark{3},
M. Salvato\altaffilmark{4},
G. Zamorani\altaffilmark{5}
T. Aldcroft\altaffilmark{6},
H. Aussel\altaffilmark{7},
H. Brunner\altaffilmark{3},
P. Capak\altaffilmark{4},
N. Cappelluti\altaffilmark{3},
F. Civano\altaffilmark{6},
A. Comastri\altaffilmark{5},
M. Elvis\altaffilmark{6},
C. Feruglio\altaffilmark{7},
A. Finoguenov\altaffilmark{3},
A. Fruscione\altaffilmark{6},
R. Gilli\altaffilmark{5},
G. Hasinger\altaffilmark{3},
A. Koekemoer\altaffilmark{8},
J. Kartaltepe\altaffilmark{9},
O. Ilbert\altaffilmark{9},
C. Impey\altaffilmark{10},
E. Le Floc'h\altaffilmark{9},
S. Lilly\altaffilmark{11}, 
V. Mainieri\altaffilmark{12},
A. Martinez-Sansigre\altaffilmark{13},
H.J. McCracken\altaffilmark{14},
N. Menci\altaffilmark{1},
A. Merloni\altaffilmark{3},
T. Miyaji\altaffilmark{15},
D.B. Sanders\altaffilmark{9},
M. Sargent\altaffilmark{13},
E. Schinnerer\altaffilmark{13},
N. Scoville\altaffilmark{4},
J. Silverman\altaffilmark{11},
V. Smolcic\altaffilmark{14},
A. Steffen\altaffilmark{4},
P. Santini\altaffilmark{1},
Y. Taniguchi\altaffilmark{16},
D. Thompson\altaffilmark{4,17},
J.R. Trump\altaffilmark{10},
C. Vignali\altaffilmark{18},
M. Urry\altaffilmark{19},
L. Yan\altaffilmark{4}
}

\altaffiltext{1}{INAF--Osservatorio Astronomico di Roma, 
via Frascati 33, Monteporzio (Rm), I00040, Italy}
\altaffiltext{2}{ASI Science data Center, via Galileo Galilei, 
00044 Frascati, Italy}
\altaffiltext{3}{Max Planck Institut f\"ur extraterrestrische Physik,
       Giessenbachstrasse 1, D--85748 Garching, Germany}
\altaffiltext{4}{California Institute of Technology, MC 105-24, 1200 East
California Boulevard, Pasadena, CA 91125} %Scoville, Capak, Thompson salvato
\altaffiltext{5}{INAF--Osservatorio Astronomico di Bologna, via Ranzani 1,
I--40127 Bologna, Italy}
\altaffiltext{6}{Harvard-Smithsonian Center for Astrophysics, 60 Garden
Street, Cambridge, MA 02138 } %elvis aldcroft fruscione civano
\altaffiltext{7}{CEA/DSM-CNRS, Universite' Paris Diderot, DAPNIA/SAp,
Orme des Merisiers, 91191, Gif-sur-Yvette, France}
\altaffiltext{8}{Space Telescope Science Institute, 3700 SanMartin
Drive, Baltimore, MD 21218} % Koekemoer
\altaffiltext{9}{Institute for Astronomy, University of  Hawaii, 
2680 Woodlawn Drive, Honolulu, HI, 96822 USA} 
\altaffiltext{10}{Steward Observatory, University of Arizona, Tucson, AZ 85721}
\altaffiltext{11}{Department of Physics, Eidgenossiche Technische Hochschule
  (ETH), CH-8093 Zurich, Switzerland}
\altaffiltext{12}{European Southern Observatory, Karl-Schwarzschild-str. 2,
  85748 Garching bei M\"unchen, Germany}
\altaffiltext{13}{Max Planck-Institut f\"ur Astronomie, K\"onigstuhl 17, D-69117 
Heidelberg, Germany}
\altaffiltext{14}{Institut d'Astrophysique de Paris, UMR7095 CNRS,
  Universit\'e Pierre \& Marie Curie, 98 bis boulevard Arago, 75014 Paris,
  France}
\altaffiltext{15}{1IA-UNAM-Ensenada, Mexico}
\altaffiltext{16}{Department of Physics, Ehime University, 2-5 Bunkyo-cho
Matsuyama 790-8577, Japan}
\altaffiltext{17}{Large Binocular Telescope Observatory, University of Arizona, 
933 N. Cherry Ave., Tucson, AZ  85721-0065,   USA} % Thompson
\altaffiltext{18}{Universita' di Bologna, via Ranzani 1, Bologna, Italy}
\altaffiltext{19}{Yale Center for Astronomy and Astrophysics, Yale University,
  P.O. Box 208121, New Haven CT 06520-8121, USA}

\email{fiore@oa-roma.inaf.it}

\begin{abstract}
\ A large population of heavily obscured, Compton-thick AGN is
predicted by AGN synthesis models for the cosmic X-ray background and
by the ``relic'' super-massive black-hole mass function measured from
local bulges. However, even the deepest X-ray surveys are inefficient
to search for these elusive AGN. Alternative selection criteria,
combining mid-infrared with near-infrared and optical photometry, have
instead been successful to pin-point a large population of Compton
thick AGN.  We take advantage of the deep {\it Chandra} and {\it
Spitzer} coverage of a large area (more than 10 times the area covered
by the {\it Chandra} deep fields, CDFs) in the COSMOS field, to extend
the search of highly obscured, Compton-thick active nuclei to higher
luminosity. These sources have low surface density and large samples
can be provided only through large area surveys, like the COSMOS
survey. We analyze the X-ray properties of COSMOS MIPS sources with
24$\mu$m fluxes higher than 550$\mu$Jy.
For the MIPS sources not directly detected in the {\it Chandra} images
we produce stacked images in soft and hard X-rays bands. To estimate
the fraction of Compton-thick AGN in the MIPS source population we
compare the observed stacked count rates and hardness ratios to those
predicted by detailed Monte Carlo simulations including both
obscured AGN and star-forming galaxies. The volume density of Compton
thick QSOs (logL(2-10keV)=44-45 ergs s$^{-1}$, or log$\lambda
L_\lambda (5.8\mu$m)=44.79-46.18 ergs s$^{-1}$ for a typical infrared
to X-ray luminosity ratio) evaluated in this way is $(4.8\pm1.1)
\times10^{-6}$ Mpc$^{-3}$ in the redshift bin 1.2--2.2.  This density
is $\sim44\%$ of that of all X-ray selected QSOs in the same redshift
and luminosity bin, and it is consistent with the expectation of
most up-to-date AGN synthesis models for the Cosmic X-ray background
(Gilli et al. 2007). The density of lower luminosity Compton-thick AGN
(logL(2-10keV)=43.5-44) at z=0.7--1.2 is $(3.7\pm1.1) \times10^{-5}$
Mpc$^{-3}$, corresponding to $\sim67\%$ of that of X-ray selected
AGN. The comparison between the fraction of infrared selected, Compton
thick AGN to the X-ray selected, unobscured and moderately obscured
AGN at high and low luminosity suggests that Compton-thick AGN follow
a luminosity dependence similar to that discovered for Compton-thin
AGN, becoming relatively rarer at high luminosities. We estimate that
the fraction of AGN (unobscured, moderately obscured and Compton
thick) to the total MIPS source population is $49\pm10\%$, a value
significantly higher than that previously estimated at similar
24$\mu$m fluxes.  We discuss how our findings can constrain AGN
feedback models.

\end{abstract}

\section{Introduction}

Understanding how galaxies formed and how they became the complex
systems we observe in the local Universe is a major theoretical and
observational effort, mainly pursued using large and deep
multi-wavelength surveys. The ubiquitous observation of 'relic'
super-massive black holes (SMBH) in the center of nearby bulge
dominated galaxies, and the discovery of tight correlations between
their masses and bulge properties suggest strong links and feedbacks
between SMBH, nuclear activity and galaxy evolution (Gebhardt et
al. 2000, Ferrarese \& Merritt 2000 and references therein).  The peak
of both nuclear (AGN) and star-formation activities is at z$\gs1$
(Boyle et al. 1988, Madau et al. 1996, Hopkins et al. 2006, Brandt \&
Hasinger 2005 and references therein), possibly due to the fact that
more gas is available at high-z for both AGN fueling and
star-formation.  In recent years, a number of evidences have been
accumulated showing similar mass-dependent evolution for galaxies and
AGN (i.e. black holes). We have robust evidence that massive galaxies
are characterized by a star formation history that peaks at z$\gs2$
(Renzini 1996), while lower mass galaxies are typically younger
systems (Cowie et al. 1996). Similarly, the density of the high
luminosity AGN (QSO hereinafter, see Table 1 for a quantitative
definition) peaks at z$\gs2$ and declines strongly afterwards, while
lower luminosity AGN follow a much smoother behavior, peaking at lower
redshifts, z=1-1.5 (Ueda et al. 2003, Cowie et al. 2003, Fiore et
al. 2003, Hasinger et al. 2005, La Franca et al. 2005). These trends
in the evolution of galaxies and AGN have been dubbed ``downsizing''
(e.g. Cowie et al. 1996, Franceschini et al. 1999), meaning that large
structures tend to have formed earlier and have grown faster than
smaller structures. The co-evolution of galaxies and AGN and their
downsizing depends on feedback between nuclear and galactic activities
(Silk \& Rees 1998, Fabian 1999, Granato et al. 2001, 2004, Menci et
al 2006, Bower et al. 2006, Daddi et al. 2007). QSOs, presumably
hosted in high mass progenitors, must somehow be more efficient at
inhibiting star-formation in their host galaxies, by heating the
interstellar matter through winds, shocks, and high energy radiation
(Silk \& Rees 1998, Fabian 1999, Granato et al. 2001, 2004, Di Matteo
et al. 2005, Menci et al. 2006, 2008, Bower et al. 2006, Li et
al. 2007). In this picture, a short, powerful AGN phase is believed to
precede the phase when a galaxy is found in a passive status with red
optical and UV colors, most of the star-formation having been
inhibited by the AGN activity. Conversely, feedback from less powerful
AGN, presumably hosted in low mass progenitors, is more effective in
self-regulating accretion and star-formation, and cold gas is left
available for both processes for a much longer time.  The same cold
gas can intercept the line of sight to the nucleus, and therefore a
natural expectation of this scenario is that the fraction of obscured
AGN to the total AGN population is large at low AGN luminosities and
decreases at high luminosities, as it is indeed observed (Lawrence \&
Elvis 1982, Ueda et al. 2003, Steffen et al. 2003, La Franca et
al. 2005, Treister \& Urry 2005, 2006, Maiolino et al. 2007, Hasinger
2008, Trump et al. 2008, Della Ceca et al. 2008).  Powerful AGN clean
their sight-lines more quickly than low luminosity AGN, and therefore
the fraction of active objects caught in an obscured phase decreases
with the luminosity. Under this hypothesis, the fraction of obscured
AGN can be viewed as a measure of the timescale over which the nuclear
feedback is at work (Menci et al. 2008). In this respect the trend of
the fraction of obscured AGN with the luminosity can be regarded as a
manifestation of the AGN downsizing. Obscured AGN are thus
laboratories in which to investigate feedback in action.

We call ``AGN'' objects with their SMBH in an active status.  This
does not imply that the AGN must dominate the bolometric luminosity,
but simply that it is possible to recognize its emission in at least
one of the bands of the electromagnetic spectrum. We call
``unobscured'' those AGN in which the optical and soft X-ray nuclear
light is not blocked by gas and dust along the line of sight. AGN in
which the nuclear light is blocked or reduced by dust and gas along
the line of sight are called ``obscured''.  We further distinguish
between moderately obscured AGN (or Compton-thin) and highly obscured
AGN (or Compton-thick, CT), see Table 1 for quantitative definitions.
While unobscured and moderately obscured AGN can be efficiently
selected in current X-ray surveys, even the deepest {\it Chandra} and
{\it XMM-Newton} surveys detected directly only a handful of CT AGN
(e.g.  Tozzi et al. 2006, Comastri 2004).  However, the SMBH mass
function obtained by integrating these X-ray luminosity functions
falls short by a factor $\sim1.5-2$ (depending on the assumed
efficiency in the conversion of gravitational energy into radiation)
of the SMBH mass function, evaluated by using the M$_{BH} - \sigma_V$
/ M$_{BH} -$M$_B$ relationships and the local bulge's luminosity
function (the 'relic' SMBH mass function, Marconi et al., 2004,
but also see Merloni \& Heinz 2008).  AGN synthesis models for the
Cosmic X-ray Background (CXB, Treister et al. 2004, Treister \& Urry
2005, Gilli et al. 2007) predict a large enough volume density of CT
AGN to reconcile the 'active' and 'relic' SMBH mass functions.

Obscured AGN, including CT ones, can be recovered, thanks to the
reprocessing of the AGN UV emission in the infrared, by selecting
sources with mid-infrared (and/or radio) AGN luminosities but faint
near-infrared and optical emission.  Houck et al. (2005), Weedman et
al. (2006a,b), Yan et al. (2007), Polletta et al. (2008) obtained
Spitzer IRS spectra of large samples of relatively bright 24$\mu$
sources (F(24$\mu) > 0.7$mJy) with faint optical counterparts, finding
that the majority are AGN dominated.  The small UV rest frame
luminosity implies significant obscuration in these objects. Indeed,
Polletta et al. (2006) used X-ray data to infer that some of these
infrared bright QSOs are Compton-thick.  Martinez-Sansigre et
al. (2005, 2007, 2008) obtained optical and Spitzer IRS spectra of
sources with F(24$\mu) > 0.3$mJy, and faint optical and near infrared
counterpars, finding that most are highly obscured, type 2 QSOs.
Brand et al. (2007) obtained infrared spectroscopy of 10 sources with
F(24$\mu) > 0.8$mJy and faint optical counterpars, finding that 6
exhibit broad H$\alpha$ lines. Since both the narrow line region and
the UV continuum are strongly extinted, they suggest that the
obscuration is due to dust on large scales, withing the host galaxies.
Dey et al. (2008) obtained optical spectroscopy of a rather large
sample of objects with extreme F($24\mu$m)/F(R) flux ratios and
F(24$\mu) > 0.3$mJy.  They found a redshift distribution centered at
z$\sim2$, implying large luminosities, and concluded that both
star-formation and nuclear activity are probably contributing to these
luminosities.  Finally, Daddi et al. (2007) and Fiore et al. (2008,
F08 hereafter) suggested that the majority of the so called `IR
excess' sources in the CDFS, with an extreme F($24\mu$m)/F(R) flux
ratios and F(24$\mu$m) as low as 40$\mu$Jy, are highly obscured,
possibly CT AGN at z=1--3. Although Donley et al. (2008) and Pope et
al. (2008) disagree with the latter two studies on the AGN fraction at
faint 24$\mu$m fluxes, it is clear that selecting bright 24$\mu$m
sources with extreme F($24\mu$m)/F(R) flux ratios may represent a
promising method to complement X-ray surveys in obtaining sizable
samples of CT AGN and so completing the census of accreting SMBH at
these redshifts.

Here we apply and extend this approach to the COSMOS field to
estimate the total (unobscured, moderately obscured and CT) AGN
fraction to the full MIPS 24$\mu$m galaxy population. We take
advantage of its deep and uniform coverage at infrared, optical and
X-ray wavelengths, to select and validate samples of CT AGN at
z=0.7--2. The COSMOS sample contains sources which have IR/optical
properties similar to those in the {\it Chandra} Deep Fields (CDFs)
but are $\sim10$ times brighter, and are therefore much more luminous
than the CDFs AGN at the same redshift. Our goal is to select a
sizable sample of high luminosity CT QSOs to measure accurately their
volume density and to understand whether their obscuration properties
are similar to those of lower luminosity AGN. This will allow us to
understand whether the correlation between the fraction of obscured
AGN and luminosity holds for CT QSOs, and to extend this study up to
z$\sim2$.  Such luminous sources are rare, and only taking advantage
of the large area covered by COSMOS they can be found in significant
number to make statistical studies. The total area covered by COSMOS
is 2 deg$^2$ but in this work we limit the analysis to the area
covered by deep Chandra observations ($\sim0.9$ deg$^2$, $\gs10$ times
the area in CDFs).

The paper is organized as follows: Section 2 presents the datasets
used in this work and the selection of the CT QSO sample from the {\it
Spitzer} MIPS 24$\mu$m COSMOS sample; Section 3 discusses the X-ray
properties of the MIPS selected sources; Section 4 presents the result
of fitting the broad band spectral energy distributions (SEDs) of the
MIPS selected sources with galaxy and AGN templates; Section 5
presents our evaluation of the infrared selected CT QSO volume density
and finally Section 6 gives our conclusions. A $H_0=70$ km s$^{-1}$
Mpc$^{-1}$, $\Omega_M$=0.3, $\Omega_{\Lambda}=0.7$ cosmology is
adopted throughout.

\begin{table}
\caption{\bf AGN definitions used in this paper}
\begin{tabular}{lcc}
\hline
\hline
AGN type & A$_V$ & logN$_H$\\
         &     & \cm2 \\
\hline
Unobscured & $<5^a$ & $\ls 22$ \\
Moderately obscured (Compton-thin) & $>5$ & 22--24 \\
Highly obscured (Compton-thick) & $\gs20^b$ & $>24$ \\
\hline
AGN type & logL(2-10keV) & logN$_H$\\
         & \cgs          & \cm2 \\
\hline
Unobscured QSO & $>44$   & $\ls 22$ \\
Compton-thin QSO & $>44$ & 22--24 \\
Compton-thick QSO & $>44$ & $>24$ \\
\hline
\end{tabular} 

$^a$ From Simpson et al. (1999);
$^b$ Allowing for the fact than obscured AGN and QSO can
have gas-to-dust ratios much smaller than the Galactic value, see
e.g. Maiolino et al. 2001 and Martinez-Sansigre et al. (2006).

\end{table}

\section{Datasets and sample selection}

\begin{figure*}[ht]
\centering
\begin{tabular}{cc}
\includegraphics[width=8cm, angle=0]{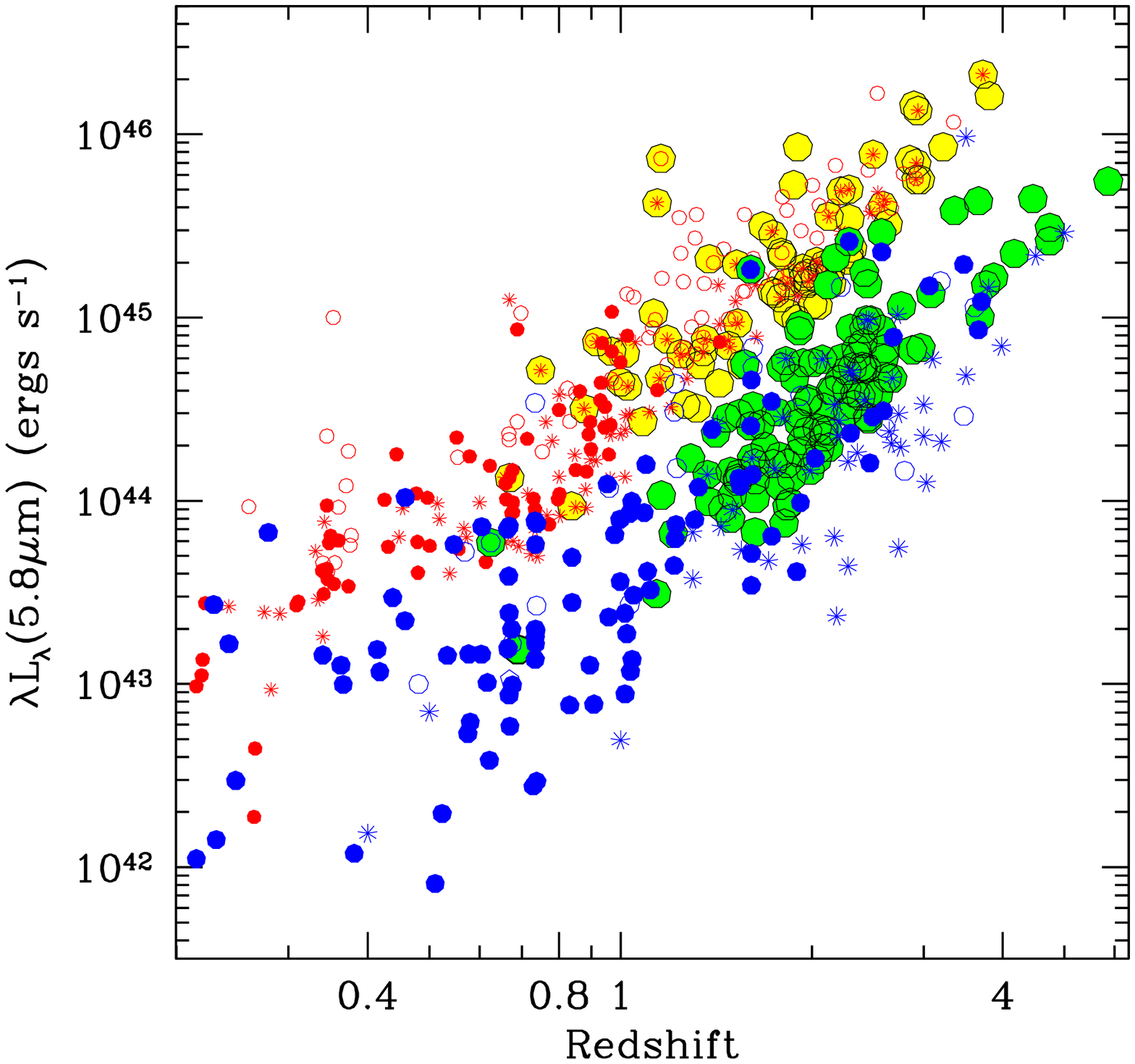}
\includegraphics[width=8cm, angle=0]{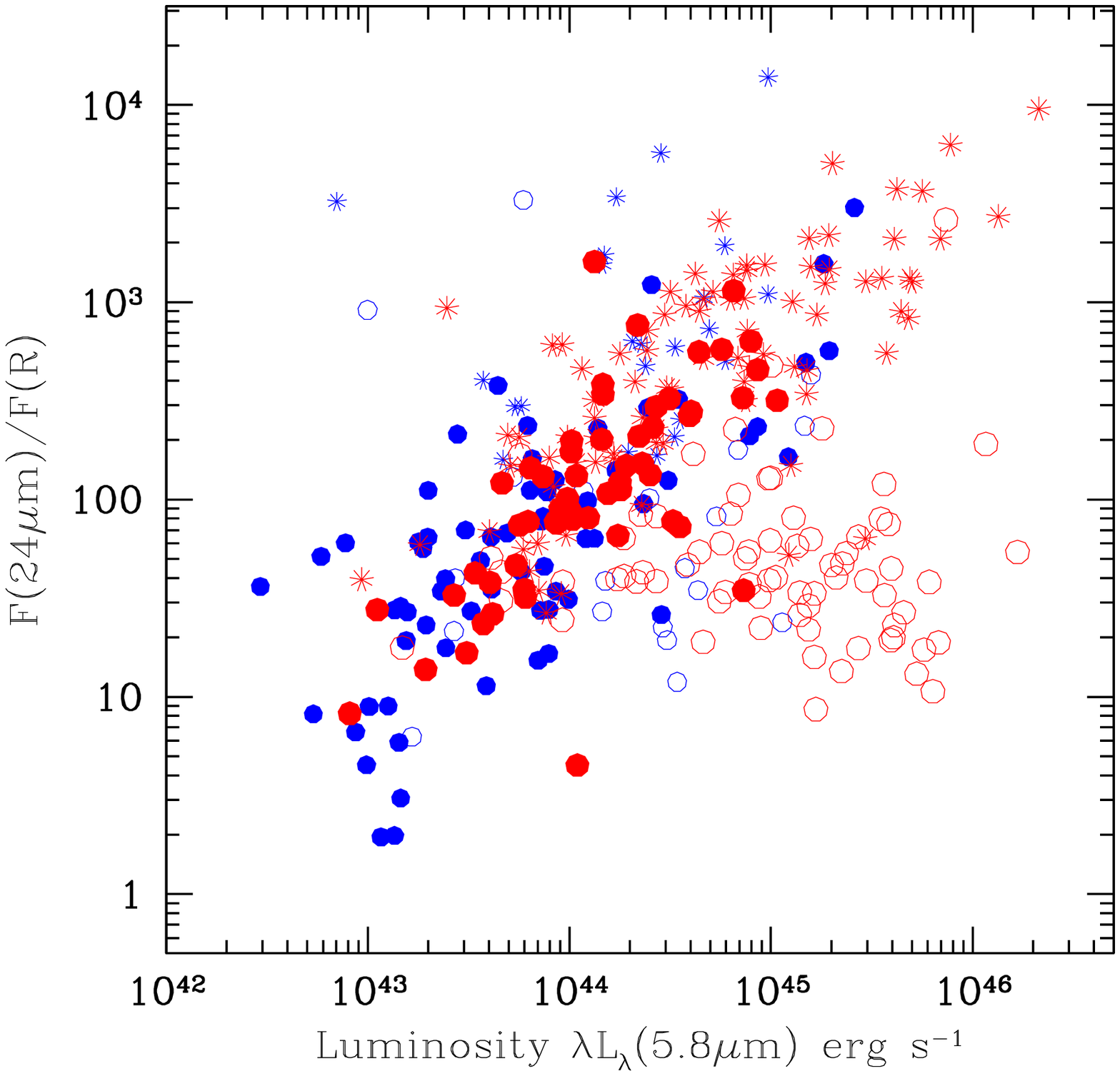}
\end{tabular}
\caption{[Left panel:] The redshift-infrared luminosity, log$(\lambda
L_\lambda (5.8\mu m))$, plane for the C-COSMOS sources (red and
yellow points) and the CDF-S sources (blue and green points). Red and
blue circles are sources directly detected in the X-rays. Yellow and
green points are 24$\mu$m, non X-ray detected sources with
$F(24\mu{\rm m})/F(R) >1000$ and $R-K >4.5$. [Right panel:]\
F($24\mu$m)/F(R) as a function of the 5.8$\mu$m luminosity for two
X-ray source samples (GOODS-MUSIC, blue symbols, C-COSMOS, red
symbols).  In both panels open circles correspond to spectroscopic
type 1 AGN, filled circles to non type 1 AGN and asterisks to objects
with photometric redshifts.  Note that F($24\mu$m)/F(R) of non broad
line AGN is strongly correlated with the luminosity at 5.8$\mu$m.  }
\label{mirolir}
\end{figure*}

\begin{figure}[h]
\centering
\includegraphics[width=8cm, angle=0]{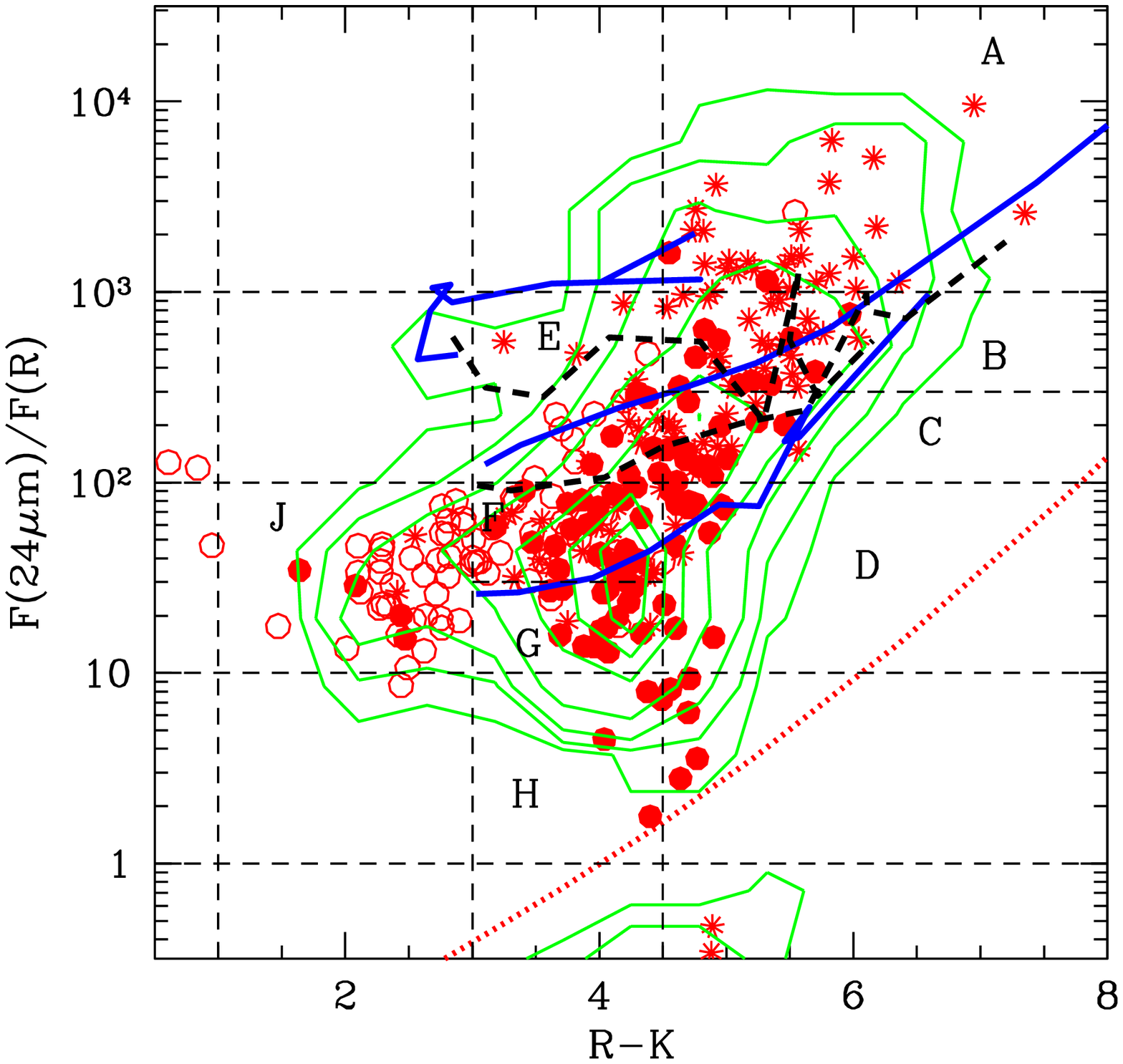}
\caption{F($24\mu$m)/F(R) as a function of R-K color for the
X-ray detected population with a luminosity $>10^{38}$ ergs s$^{-1}$.
Open circles = type 1 AGN; filled circles = non type 1 AGN; asterisks
= photometric redshifts.  Iso-density contours of all COSMOS 24$\mu$m
sources are overlaid to the plot. Expected evolution from z$=0$ to
z=$5$ for three obscured AGN (blue continuous lines), two star-burst
(black dashed lines) and a passive elliptical galaxy (red dotted line)
are shown for reference (see F08 for details on the SEDs used). The
nine cells defined in the F($24\mu$m)/F(R) plane for the stacking
analysis are labeled with letters from A to J (see Section 2.2 for
details).}
\label{mirormk}
\end{figure}

The Cosmic Evolution Survey (COSMOS) field (Scoville et al. 2007) is a
so far unique area for its deep \& wide comprehensive multiwavelength
coverage, from the optical band with {\em Hubble} and ground based 8m
class telescopes, to infrared with {\em Spitzer}, to X-rays with {\it
XMM-Newton} and {\em Chandra}, to the radio with the VLA. The COSMOS
field samples a volume at high redshift ($z\ls2$) which is $\sim15\%$
of that sampled by SDSS at z$\ls0.15$.

For this work we use the {\em Spitzer} MIPS 24$\mu$m COSMOS catalog
derived from Cycle 2 shallow observations (Sanders et al. 2007). 
We do not use in this work the cycle
2 deep MIPS test area because it covers only a small fraction of the
{\it Chandra}-COSMOS area.  The cycle 2 catalog has been cleaned of
spurious sources (mostly asteroids) by comparison with the much deeper
MIPS mosaic obtained in Cycle 3 (Aussel et al. 2008, a reliable
catalog from this mosaic is in preparation and will be used in a
follow-up publication). We consider only MIPS 24$\mu$m sources with a
signal to noise ratio $>4$, implying a 24$\mu$m flux limit of $\sim550
\mu$Jy.  We limit the analysis to the area covered by {\it Chandra}
observations ($\sim0.9$ deg$^2$, centered on RA=10 00 20 dec=+02 09
28, Elvis et al. 2008). The sample includes 919 sources, and we refer
in the following to this sample as the COSMOS Bright MIPS Sample, see
Table 2 for further detail.

The {\em Spitzer} MIPS catalog has been cross correlated with the IRAC
catalog
\footnote{http://irsa.ipac.caltech.edu/cgi-bin/Gator/nph-scansubmit=Select\&projshort=COSMOS}
(Sanders et al. 2007) the optical multi-band catalog (Capak et
al. 2007) and the K band catalog (McCracken et al. 2008).  IRAC,
K-band and optical counterparts of the MIPS sources have been
carefully identified with a 2 step approach: first, each MIPS source
has been associated with the most likely IRAC counterpart within 2
arcsec from the MIPS centroid, then IRAC positions have been
correlated with the optical and K-band catalogs with a matching radius
of 0.5 arcsec.  859 sources of the COSMOS bright MIPS catalog have
counterparts in all IRAC, optical and K bands, 53 sources have
counterparts in only one or two catalogs. Finally, 4 sources have
optical counterpart too close to a bright sources (and therefore no
reliable photometry is available) while for the remaining 3 sources no
optical counterpart was assigned (they are either residual asteroids
or too faint to be detected).

The X-ray properties of the MIPS selected sources have been studied
using the {\it Chandra} data. Chandra observed the COSMOS field for a
total of 1.8 Msec.  The survey uses a series of 36 heavily overlapped
ACIS-I 50~ksec pointings to give an unprecedented uniform effective
exposure of 185~ksec over a large area. Particular care was taken in
performing accurate astrometric corrections and in the reduction of
the internal background (see Elvis et al. 2008 for details). The
residual background is very stable over the full field at $\ls 2$
counts/200~ksec over an area of 2 arcsec radius, comparable with the
{\em Chandra} beam size, see Table 2 for further details.  
Uniform coverage and low background make the {\it
Chandra} COSMOS (C-COSMOS) dataset ideal for stacking analyses.

\begin{table*}
\caption{\bf Main datasets used in this paper}
\begin{tabular}{lcccccc}
\hline
\hline
Dataset & Area & Total exposure & Typical exposure & FWHM & Detected sources & Flux limit \\
\hline
C-COSMOS  & 0.9deg$^2$ & 1.8Msec & 90-185ksec$^b$& $2''$$^c$ & 1760 & $2\times10^{-16}$$^d$\\
S-COSMOS (MIPS) & 0.9deg$^2$ & 58.2hr & 80sec & $5''$$^e$ & 919 & 550$\mu$Jy \\
\hline
\end{tabular} 

$^a$ Typical exposure; $^b$ $\sim$half of the total area is covered
with an effective exposure of $\sim185$ksec, the remaining half has an effective
exposure of $\sim90$ksec; $^c$ average for four overlapping fields; 
$^d$ \cgs 0.5-2 keV; $^e$PSF core.
\end{table*}

\subsection{Redshifts}

For 96 \% of the COSMOS bright MIPS sample either spectroscopic or 
robust photometric redshifts have been obtained. Accurate
spectroscopic redshifts are present for 394 MIPS sources, 43\% of the
sample (Lilly et al. 2007, Trump et al. 2007,2008).

Regarding photometric redshifts, we used both the computation of
Ilbert et al. (2008) and of Salvato et al. (2008), which made use of
about 30 photometric data points (including 12 intermediate filters
observed with SUBARU).  The first concentrates on objects with
I(AB)$<26.5$, with SEDs dominated by the integrated stellar population
at $\lambda <5.5 \mu$m.  The achieved accuracy for the MIPS selected
sample is $\sigma(\Delta z/(1+z))=0.01$ for I$<24$. The accuracy
degrades at I$>24$ ($\sigma(\Delta z/(1+z))=0.05$).  The second
photometric redshift catalog deals with the XMM-COSMOS (Hasinger et
al. 2007) sources that are dominated by AGN emission.  For better
results i) correction for variability, ii) luminosity priors for
point-like sources, and iii) a new set of SED templates have been
adopted.  Thus, for the first time an accuracy comparable to those of
photometric redshift for non-active galaxies has been achieved
($\sigma(\Delta z/(1+z))<0.02$ and less that 5\% outliers). This
second photometric dataset, as specific for X-ray detected sources,
was used when the MIPS selected source was associated to an {\it
XMM-Newton} detection. The median redshift of the COSMOS bright MIPS
sample is 0.64 with interquartile 0.36. 230 sources have z$>1$ and 56
z$>2$.  Tables 3 and 4 also give the median redshift and its
interquatile of MIPS samples selected in intervals of F(24$\mu$m)/F(R)
and R-K color, see next section.

\subsection{Optical, near infrared and mid infrared color selection}

F08 proposed a criterion, based on high mid-infrared to optical flux
ratios and red optical colors, to efficiently select candidate CT AGN,
and applied this method to the CDF-S area.  Briefly, they selected a
sample of candidate CT AGN by using the combination of
F(24$\mu$m)/F(R)$>1000$ and R-K$>4.5$ colors.  They demonstrated that
the selected sample was indeed mainly made by CT AGN through a
stacking analysis of the {\it Chandra} X-ray data. The CDF-S infrared
selected, CT AGN have z$\sim$1--3, and $\lambda L_\lambda
(5.8\mu$m)$\approx 10^{44-45}$ ergs s$^{-1}$, corresponding to
intrinsic X-ray (2-10 keV) luminosities $\approx 10^{43-44}$ ergs
s$^{-1}$. F08 limited their analysis to MIPS sources with
F(24$\mu$m)/F(R)$>1000$ and R-K$>4.5$, a minority of the full MIPS
source population. For this reason they did not compute AGN volume
densities nor AGN fractions with respect to the full CDFS MIPS galaxy
sample.  Here we apply the F08 method to the C-COSMOS field and
extend their analysis to the full MIPS source population. This allows
us to compute proper AGN volume densities and AGN fractions with
respect to the full MIPS source population.

Fig. \ref{mirolir} (left panel) compares the $5.8\mu$m luminosities
and redshifts of the CDF-S (green) and C-COSMOS (yellow) MIPS selected
sources not directly detected in X-rays and with
F(24$\mu$m)/F(R)$>1000$ and R-K$>4.5$, to those of the X-ray detected
population (red and blue symbols for the C-COSMOS and CDF-S fields
respectively). Monochromatic luminosities were computed using a linear
interpolation between the observed 8$\mu$m and 24$\mu$m fluxes at the
wavelength corresponding to 5.8$\mu$m at the source rest frame.  

\begin{table*}[t!]
\caption{\bf MIPS sources with a direct {\it Chandra} detection}
{\tiny
\begin{tabular}{lcccccccc}
\hline
\hline
Cell & F(24$\mu$m)/F(R) & R-K & \# of MIPS  & \# of X-ray det. & 
$<z>$ & $<logL(5.8\mu m)>$ & $<$logL(2-10keV)$>$ & $<$(H-S)/(H+S)$>^a$ \\
  &            &         & sources    &   &           & ergs s$^{-1}$ 
& ergs s$^{-1}$ &   \\
\hline
A & $>1000$    & $>4.5$  & 73  & 31 (40\%) & 1.55$\pm$0.53 & 45.20$\pm$0.44 
& 43.51$\pm$0.40 & 0.50$\pm$ 0.34 \\			 
B & $300-1000$ & $>4.5$  & 98  & 34 (35\%) & 0.96$\pm$0.13 & 44.58$\pm$0.27 
& 43.15$\pm$0.41 & 0.44$\pm$0.31 \\			 
C & $100-300 $ & $>4.5$  & 105 & 24 (23\%) & 0.73$\pm$0.12 & 44.12$\pm$0.19 
& 42.69$\pm$0.57 & 0.58$\pm$0.22 \\			 
D & $10-100$   & $>4.5$  & 65  & 14 (22\%) & 0.50$\pm$0.25 & 43.80$\pm$0.20 
& 42.01$\pm$0.65 & 0.48$\pm$0.28 \\			 
\hline						 
E & $100-1000$ & $3-4.5$ & 72  & 26 (36\%) & 0.90$\pm$0.37 & 44.61$\pm$0.46 
& 43.17$\pm$0.78 & 0.29$\pm$0.31 \\			 
F & $30-100$   & $3-4.5$ & 187 & 61 (33\%) & 0.48$\pm$0.21 & 43.94$\pm$0.35 
& 42.60$\pm$0.87 & 0.00$\pm$0.22\\			 
G & $10-30$    & $3-4.5$ & 144 & 19 (13\%) & 0.17$\pm$0.07 & 43.29$\pm$0.26 
& 40.85$\pm$0.85 & 0.28$\pm$0.41\\			 
H & $1-10$     & $3-4.5$ &  14 & 4  (29\%) & 0.13$\pm$0.18 & 43.16$\pm$0.68 
& 41.00$\pm$0.74 & 0.35$\pm$0.22  \\			 
\hline						 
J & $10-100$   & $1-3$   & 52  & 41 (79\%) & 1.48$\pm$0.33 & 45.30$\pm$0.32 
& 44.15$\pm$0.34 & -0.06$\pm$0.12\\
\hline
\end{tabular}

$^a$ H=1.5-6 keV, S=0.3-1.5 keV.
}

\end{table*}

\begin{table*}
\caption{\bf MIPS sources without a direct {\it Chandra} detection}
{\tiny
\begin{tabular}{lccccccccc}
\hline
\hline
Cell   & F(24$\mu$m)/F(R) & R-K & \# of MIPS & $<z>$ & $<logL(5.8\mu m)>$  
& Counts   & Counts       &  $<$(H-S)/(H+S)$>^a$ & CT AGN \\
       &                  &     & sources    &       & ergs s$^{-1}$
& 1.5-6keV & 0.3-1.5keV   &         & fraction \\
\hline
A & $>1000$     & $>4.5$   & 42  & 1.90$\pm$0.40 & 45.24 0.24 & 58.9$\pm$8.5 
& 18.0$\pm$4.7 & 0.53$\pm$0.14 & 0.94$^{+0.06}_{-0.08}$ \\
B & $300-1000$  & $>4.5$   & 64  & 1.01$\pm$0.31 & 44.37 0.40 & 54.1$\pm$8.1 
& 30.4$\pm$6.1 & 0.28$\pm$0.12 & 0.72$^{+0.14}_{-0.19}$\\
C & $100-300 $  & $>4.5$   & 81  & 0.79$\pm$0.14 & 44.01 0.17 & 42.2$\pm$7.2 
& 36.9$\pm$6.7 & 0.07$\pm$0.12 & 0.51$^{+0.17}_{-0.18}$\\
D & $10-100$    & $>4.5$   & 51  & 0.37$\pm$0.12 & 43.76 0.24 & 31.7$\pm$6.2 
& 31.1$\pm$6.2 & 0.01$\pm$0.14 & 0.42$^{+0.16}_{-0.23}$\\
\hline				      
E & $100-1000$  & $3-4.5$  & 46  & 0.70$\pm$0.22 & 44.06 0.43 & 27.7$\pm$5.8 
& 36.1$\pm$6.6 & -0.13$\pm$0.14 & 0.26$^{+0.29}_{-0.26}$\\
F & $30-100$    & $3-4.5$  & 126 & 0.37$\pm$0.11 & 43.61 0.16 & 91.6$\pm$10.6 
& 112.6$\pm$11.7 & -0.10$\pm$0.08 & 0.31$^{+0.19}_{-0.16}$\\
G & $10-30$     & $3-4.5$  & 125 & 0.23$\pm$0.06 & 43.30 0.21 & 54.8$\pm$8.2 
& 108.9$\pm$11.5 & -0.33$\pm$0.09 & $<0.40$ \\
H & $1-10$      & $3-4.5$  &  10 & 0.11$\pm$0.04 & 42.57 0.29 & 1.3$\pm$1.3 
& 10.4$\pm$3.6 & -0.78$^{+0.41}_{-0.22}$ & $<0.20$\\
\hline				      
J & $10-100$    & $1-3$    &  11 & 0.34$\pm$0.12 & 43.52 0.35 & 0.8$\pm$1.0 
& 17.7$\pm$4.6 & -0.92$^{+0.35}_{-0.08}$ & $<0.05$\\
\hline
\end{tabular}

$^a$ H=1.5-6 keV, S=0.3-1.5 keV.
}

\end{table*}

\begin{figure*}[h!]
\centering
\includegraphics[width=16cm, angle=0]{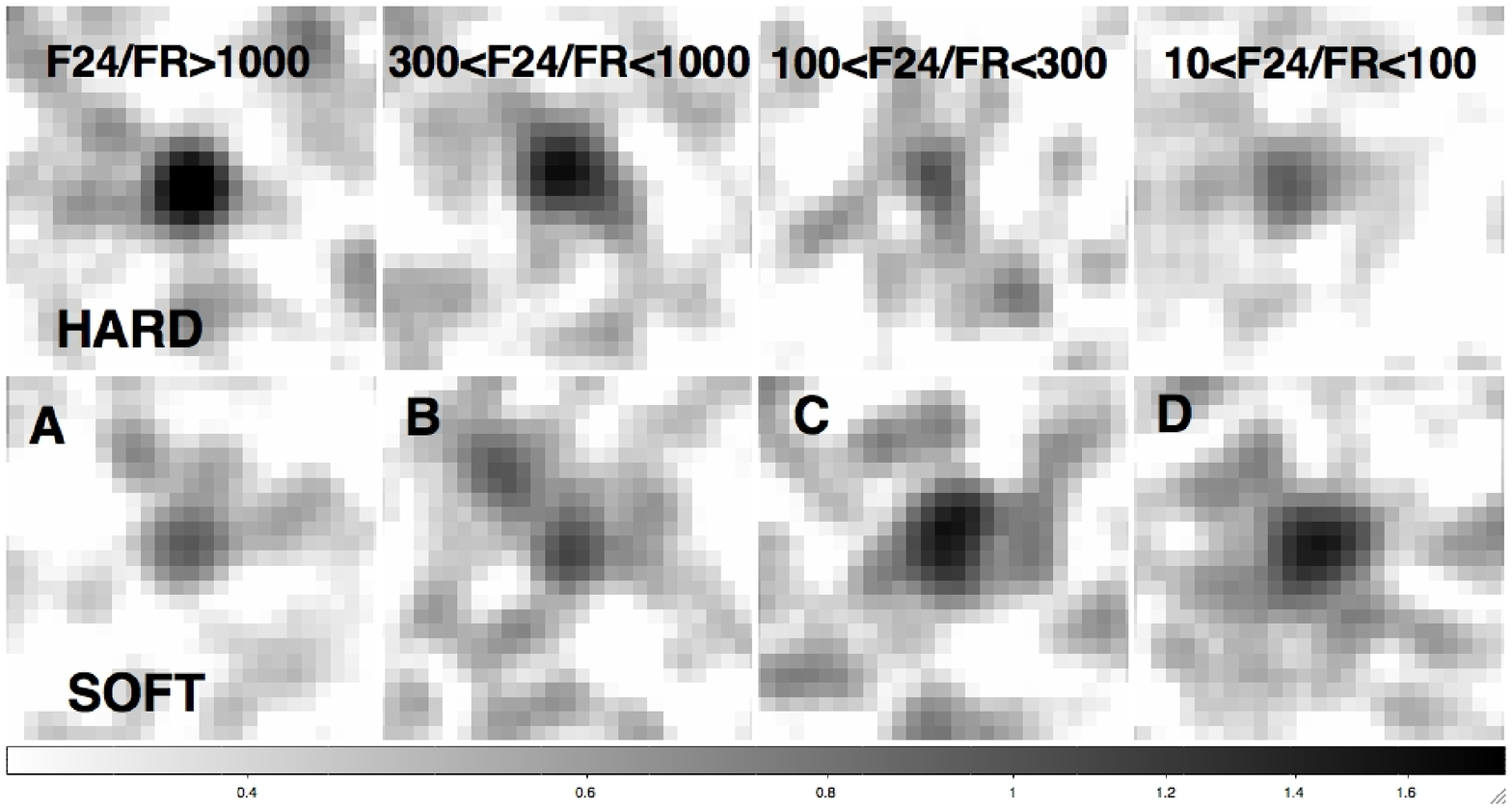}
\includegraphics[width=16cm, angle=0]{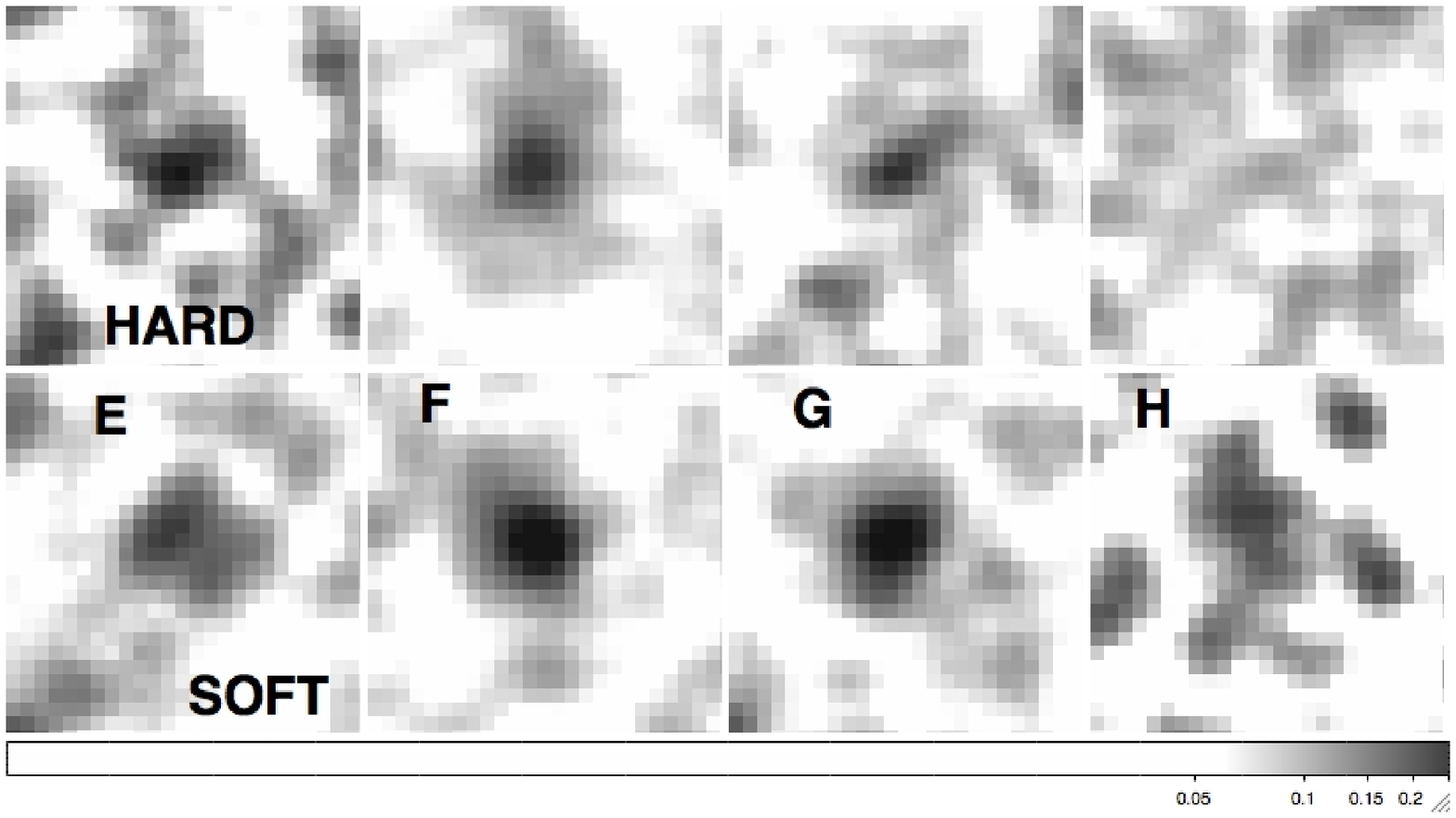}
\caption{Stacked {\it Chandra} images 
in the hard 1.5-6 keV and soft 0.3-1.5 keV bands of COSMOS MIPS sources 
not directly detected in X-rays in eight $F(24\mu{\rm
m})/F(R)$ cells (A,B,C,D,E,F,G,H). Images have sides of 12 arcsec and have
been smoothed with a gaussian with 1.5 arcsec $\sigma$. 
}
\label{8ima}
\end{figure*}

Both MIPS selected and X-ray selected COSMOS sources have luminosities
$\sim10$ times higher than the CDF-S sources at the same redshifts, as
expected since luminous QSOs are rarer than low luminosity AGN and the
C-COSMOS survey covers an area about 10 times larger than the CDF-S at
a 10 times brighter 24$\mu$m flux limit ($\sim550\mu$Jy and 40$\mu$Jy
for C-COSMOS and CDFS respectively).

Fig. \ref{mirolir} (right panel) shows F(24$\mu$m)/F(R) as a function
of the 5.8$\mu$m rest frame luminosity $\lambda L_\lambda (5.8\mu$m),
for the C-COSMOS and CDF-S X-ray sources.  Unobscured AGN (open
symbols in Fig. \ref{mirolir}) have F(24$\mu$m)/F(R) in the range
10-200, uncorrelated with $\lambda L_\lambda (5.8\mu$m), as expected
because the nuclear emission dominates both optical and mid infrared
wavelengths. Obscured AGN (filled symbols) have F(24$\mu$m)/F(R)
spanning a broader range, and correlated with $\lambda L_\lambda
(5.8\mu$m).  For a flux limited sample the luminosity is strongly
correlated with the redshift (Fig. \ref{mirolir} left
panel). Therefore, sources with high F(24$\mu$m)/F(R) (and high
5.8$\mu$m luminosity) have also redshift systematically higher than
sources with lower F(24$\mu$m)/F(R).

Fig. \ref{mirormk} shows the distribution of F(24$\mu$m)/F(R) as a
function of the R-K color.  F(24$\mu$m)/F(R) of X-ray selected,
obscured AGN is correlated with the R-K color, as found by F08 for the
CDF-S X-ray sources. The isodensity contours of the 24$\mu$m selected
sources (green curves) follow roughly the same correlation.  Unlike
the 24$\mu$m selected GOODS-MUSIC sources, we do not find a bimodal
distribution in R-K colors for the COSMOS sources at high
F(24$\mu$m)/F(R) values (see Fig. 3, right panel in F08).  This is
probably due to the higher COSMOS flux limits. The faint, blue,
star-forming galaxies found in the GOOD-MUSIC sample are not common in
the COSMOS Bright MIPS sample.

CT AGN can also have values of the F(24$\mu$m)/F(R) and R-K colors
smaller than those adopted by F08, although their fraction to the full
infrared selected population is probably small.  To properly account
for these sources we extend the F08 approach, by analyzing the full
F(24$\mu$m)/F(R) -- R-K diagram.  We divided this plane in nine cells
(see Fig. \ref{mirormk}). The boundaries of the cells were chosen
according to the following criteria: 1) cover most of the
F(24$\mu$m)/F(R) -- R-K plane; 2) sample regions not to big in this
plane; 3) but at the same time regions containing a number of sources
big enough to be statistically meaningful.  These nine cells contain
most of the COSMOS Bright MIPS sample (87\%). Our goal is to estimate
the fraction of CT AGN to the total MIPS source population in each of
these cells.  Table 3 and 4 give the number of MIPS selected sources
in each cell, along with the median redshift and luminosity (with
their interquartile ranges) of both MIPS sources with (Table 3) and
without (Table 4) a direct X-ray detection.

\section{X-ray properties of the 24$\mu$m selected sources}

\subsection{Sources with a direct X-ray detection}

232 sources of the COSMOS Bright MIPS sample are present in the
C-COSMOS and XMM-COSMOS catalogs (Elvis et al. 2008, Civano et
al. 2008, Brusa et al. 2007, 2008a). In addition to these sources we
also consider as ``detections'' 47 sources with more than 10
background-subtracted {\it Chandra} counts in the full 0.5-7 keV band,
within 5 arcsec of the position of the MIPS source but not present in
the C-COSMOS and XMM-COSMOS catalogs. This allows us to identify: 1)
faint X-ray sources (given the average background, 10
background-subtracted counts in a 5 arcsec radius area corresponds to
a probability $\sim 5\times 10^{-3}$ that the detected counts are due
to a background fluctuation); 2) MIPS sources with a nearby X-ray
source; and 3) MIPS sources found in X-ray groups and clusters of
galaxies. The total number of MIPS sources with an X-ray counterpart
is 279 ($\sim 30\%$ of the full MIPS sample, 254 sources in the nine
cells defined in Fig. \ref{mirormk}). 11 sources do not have K band
detections, while the remaining 14 sources are scattered in the
diagram outside the considered cells (R-K$<$1 and
F(24$\mu$m)/F(R)$<10$). Table 3 gives the fraction of MIPS sources
with a direct {\it Chandra} and/or {\it XMM-Newton} detection in the 9
F(24$\mu$m)/F(R) and R-K cells. The fraction of MIPS sources with a
direct X-ray detection is minimum ($\sim13$\%) in cell G, while it is
maximum (79\%) in cell J. Most of the X-ray sources in the latter cell
have been spectroscopically identified (95\%) and the majority turned
out to be type 1 QSOs (90\%), as expected from their colors.

Table 3 gives the median and interquartile range of the X-ray
and infrared luminosities of the MIPS sources with X-ray detection in
the 9 cells. The X-ray luminosity is computed in the rest frame 2-10
keV band to ease the comparison with previous studies. It is computed
from the observed 0.5-7 keV flux, assuming a power law spectrum 
$F(E)\propto E^{-alpha_E}$ with
energy index $\alpha_E=0.8$ and Galactic N$_H$. It is not corrected
for absorption, and therefore it should be considered a lower limit to
the intrinsic luminosity. The log ratio between the observed 5.8$\mu$m
and 2-10 keV median luminosities of the 35 type 1 AGN in cell J is
1.15. This can be considered to be little affected by absorption and
therefore representative of the ratio between the intrinsic AGN
luminosities.  The highest luminosity log ratio is for the objects in
bin A. Its value (1.7) is significantly higher than in bin J, thus
suggesting some obscuration or intrinsically low X-ray emission in the
X-ray counterparts of the MIPS sources in this cell.

Table 3 also gives the median and interquartile of the hardness ratios
(H-S)/(H+S) computed using the source counts detected in the 0.3-1.5
keV (S) and 1.5-6 keV (H) bands.  The hardness ratio expected for a
power law spectrum with $\alpha_E=0.8$ and no absorption in addition
to the Galactic one is -0.1. Values between 0.2 and 0.8 imply
absorbing column densities between $10^{22}$ and $10^{23}$ cm$^{-2}$
(i.e. Compton-thin absorbers) at typical redshifts of 1-2. Note that
the hardness ratio of X-ray sources with R-K$>$4.5 is systematically
higher than that of bluer sources. Similar results were found by
Mignoli et al. (2004), Brusa et al. (2005) and Mainieri et
al. (2007). Indeed, the sample of R-K$>$4.5 sources is dominated by
narrow line AGN (only 1 broad line AGN out of 43 objects with narrow
line or absorption line optical spectra, 30 of which with
logL(2-10keV)$>42$ ergs s$^{-1}$, and therefore likely to host an
active nucleus).  Converely, 69 of the 122 objects with R-K$<$4.5 and
logL(2-10keV)$>42$ ergs s$^{-1}$ are broad line AGN.

\subsection{X-ray stacking analysis of the sources without a 
direct X-ray detection}

The total number of COSMOS bright MIPS sources without a direct X-ray
detection is 640.  Twenty-three sources are bright stars while 610
sources have either a spectroscopic or a photometric redshift (224 and
386, respectively).  Finally, seven sources are either strongly
blended (4) or too faint in any optical band; in both cases the
photometry is poor and unreliable and photometric redshift are not
available. Table 4 gives the number of sources without a direct X-ray
detection in each of the 9 F(24$\mu$m)/F(R) and R-K cells along with
their median infrared luminosities and interquartile ranges. The
median redshifts and 5.8$\mu$m luminosities of the MIPS sources
without a direct X-ray detection in all cells but J are similar to
those with X-ray detection. This suggests that most if not all QSOs in
cell J have been detected in X-rays, and that the remaining sources
are inactive galaxies or very faint AGN.

To gain information on the X-ray properties of the MIPS sources
without a direct X-ray detection we performed a detailed stacking
analysis of the {\it Chandra} counts at the position of all MIPS
source for the samples in the 9 F(24$\mu$m)/F(R) and R-K
cells. Indeed, the {\it Chandra} deep and uniform coverage allows us
to increase the sensitivity (by factors of $\times 10-100$) using the
stacking technique (Daddi et al. 2007, Steffen et al. 2007, F08).
Stacking analysis was performed using both the web-based tool for
stacking analysis of {\it Chandra} data prepared by T. Miyaji
(http://saturn.phys.cmu.edu/cstack/) and software developed at ASDC by
S. Puccetti. Results from the two software in the standard bands
0.5-2 keV and 2-8 keV were fully consistent for all 9
samples. Stacking was performed on single ACIS-I exposures and on the
combined mosaic. The results were again fully consistent.

Errors on stacked net source counts and count rates are computed by
using both Poisson statistics and a ``bootstrap'' method (by
resampling the objects in the input source list). 500 bootstrapped
stacked count rate are generated for each source list.  Poisson errors
turned out to be smaller than the bootstrap error by 3-5\%. In the
following analysis we conservatively increase the Poisson error by
10\% to account for other possible systematic errors.

Stacking was performed in two energy band, to allow the evaluation of
a hardness ratio.  The signal to noise at high X-ray energies is
limited by the internal background, whose spectrum is nearly constant as a
function of the energy. Indeed the 2-8 keV band has an internal
background $\sim 4$ times higher than the 0.5-2 keV band. To optimize
the analysis at high energies we performed the stacking in several
different energy bands and chose the band which gives the highest
signal to noise ratio for most of the nine samples. The 1.5-6 keV band gave
the best results in terms of signal to noise.  In the following we
present hardness ratio computed by using this band and the 0.3-1.5 keV
band. We performed similar analysis using the 1.5-4 keV and 1.5-5 keV
bands as the higher energy band obtaining always qualitatively similar
results.

Source extraction regions were also chosen to optimize the signal to
noise ratio.  We adopted a box with 5 arcsec side (100 ACIS-I square
pixels area). A slightly lower signal to noise ratio is obtained using
boxes with 3 and 6 arcsec side. 

\begin{figure}
\centering
\includegraphics[width=8cm, angle=0]{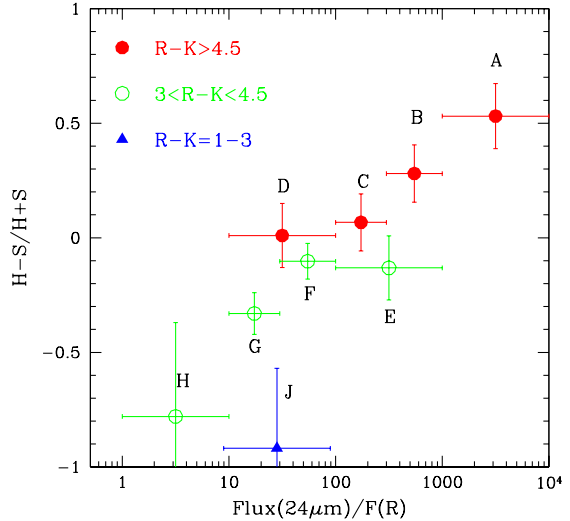}
\caption{Average hardness ratio (H-S)/(H+S) as a function of
F(24$\mu$m)/F(R) for the sources without direct X-ray detection.
Filled circle = cells A, B, C , D; open circles = cells E, F, G, H;
filled triangle = cell J. }
\label{hrmiroc}
\end{figure}

Stacked counts in both energy bands are given in Table 4. Detections
with a signal to noise higher than 4 are obtained in all cells but
cells H and J.  Fig. \ref{8ima} shows the stacked images of the
sources in all cell but J and in the two energy bands, while Table 4
and Fig. \ref{hrmiroc} shows the hardness ratio (H-S)/(H+S) derived
from the stacking analysis as a function of F(24$\mu$m)/F(R) for the 9
cells.  The samples with the highest $F(24\mu$m)/F(R) and red colors
(cells A and B) also have the hardest X-ray hardness ratios. A common
concern in stacking analyses is that the results may not be
representative of population properties, if they are biased by one or
a few sources in the stack. We detect in the stack of the sources in
cells A and B 59 and 54 background subtracted counts in the 1.5-6 keV
band, respectively.  Sources can enter in the stack only if they give
less than 10 counts in a region of 5 arcsec radius around the MIPS
position in the full 0.5-7 keV band.  Sources with an higher number of
counts are excluded. Given this threshold the stacked counts in 
cells A and B must be produced by at least 6 sources, and probably
by many more, since it would be highly unlikely to have 6 sources near
the chosen threshold and all the rest with zero counts.

The hardness ratios of the stacks of the sources without a direct
X-ray detection in cells A and B are similar to the median hardness
ratios of the X-ray detected sources in the same cells (Table
3). Taken at face values, these hardness ratios can be explained by a
power law spectra with energy index $\sim-0.5$ and $\sim0$
respectively, reduced at low energy by Galactic absorption only. Since
neither AGN or known star-forming galaxies have such hard emission
spectrum, the observed hardness ratio strongly suggest significant
rest frame obscuration.  However, for the stacked samples it is not
easy to convert a hardness ratio in a typical column density. A direct
conversion could produce contradictory results because the X-ray
spectrum of the MIPS sources is likely to be more complex than a
simple obscured power law. For example, a scattering component can
dominate the soft X-ray counts. Furthermore, samples corresponding to
different cells have different average redshifts, and therefore
comparing hardness ratios is not straightforward, as they are biased
towards measuring 'softer' HRs for an obscured AGN at high
redshift. To investigate further these issues we performed detailed
Monte Carlo simulations as described in the following section.

\subsection{Simulations to assess the fraction of obscured AGN in the 
24$\mu$m source samples}

Following F08, we used the observed flux in the stacked images,
together with the hardness ratio H-S/H+S, to constrain the fraction of
CT AGN in the MIPS samples.  For the MIPS sources in each of the 9
F(24$\mu$m)/F(R) and R-K cells we generated simulated X-ray count
rates and hardness ratios as a function of the fraction of obscured
AGN to the total MIPS source population in each cell, assuming
that the sources without an X-ray detection are either obscured AGN or
star-forming galaxies. We started from the observed redshift and
infrared luminosities. For the AGN we assumed the log($\lambda
L_\lambda (5.8\mu$m)/L(2-10 keV)) luminosity ratio given by the
following relationship to obtain unobscured 2-10 keV luminosities:

$$logL(2-10keV)=43.574+0.72(logL(5.8\mu m)-44.2)$$\\ 
$$if\  logL(5.8\mu m)>43.04 \eqno{1}$$\\
$$logL(2-10keV)=logL(5.8\mu m)-0.3$$\\  
$$if\  logL(5.8\mu m)<43.04$$

\begin{figure}[h]
\centering
\includegraphics[width=8cm]{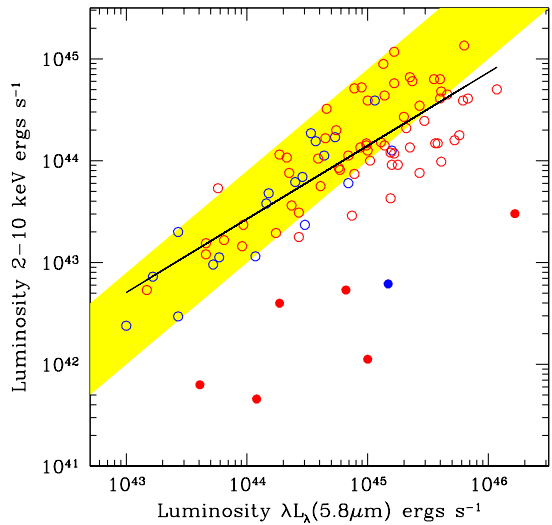}
\caption{The 2-10 keV luminosity (not corrected for obscuration) as a
function of the 5.8$\mu$m luminosity for 2 samples of type 1 AGN:
C-COSMOS (red symbols) and CDF-S (blue symbols). For most of the
points this 2-10 keV luminosity can be considered representative of
the intrinsic X-ray luminosity. Filled circles identify seven outliers
with small X-ray luminosity with respect to their infrared luminosity,
suggesting X-ray absorption in these sources, and therefore that the
intrinsic X-ray luminosity is likely under-estimated in these
cases. For this reason these sources have been excluded from the
analysis. The black solid line is the best fit linear regression in
eq.1.  The shaded region is the linear extrapolation of the intrinsic
X-ray-to-mid-infrared luminosity ratio found in the local Universe
(Lutz et al. 2004).}
\label{lxl58}
\end{figure}

This has been calibrated using the Type 1 AGN in the CDF-S (Brusa et
al. 2008b) and C-COSMOS (Civano et al. 2008) fields as shown in
Fig. \ref{lxl58}.  The derived relation assumes that the 2-10 keV
luminosity, computed directly from the observed fluxes without any
correction for intervening absorption, can be considered
representative of the intrinsic X-ray luminosity.  This is a good
approximation for most of the points. However, seven outliers (one in
the CDFS sample and six in the C-COSMOS sample), show a low X-ray
luminosity, given their infrared luminosity, suggestive of significant
X-ray obscuration in these sources. These seven points were therefore
excluded from the analysis.  The logL(2-10keV)--log($\lambda L_\lambda
(5.8\mu$m) linear regression coefficient in eq. 1 turns out to be very
similar to that found by Steffen et al. (2006, 0.643).  The
relationship in eq. 1 is consistent with the log($\lambda L_\lambda
(5.8\mu$m)/L(2-10 keV)) luminosity ratio of the highly obscured AGN in
F08, and Silva et al. (2004).  In particular, it is well consistent
with the ratio found for the type 2 QSO IRAS09104+4109 (Piconcelli et
al. 2007, 2008).  For the star-forming galaxies we assumed a
log($\lambda L_\lambda (5.8\mu$m)/L(2-10 keV)) luminosity ratio of
2.38 with a gaussian dispersion of 0.2 (see Ranalli et al. 2003).

We assumed that the star-forming galaxies are not obscured in X-rays,
while the AGN are highly obscured. We used the Gilli et al. (2007)
N$_H$ distribution. This distribution is rapidly increasing towards
high column densities, it peaks at about $10^{23.5}$ and then slowly
decreases in the Compton-thick regime.  The shape is obtained by
requiring a simultaneous fit of both the X-ray background spectrum and
the source counts in different energy bands (0.5-2 keV , 2-10 keV and
5-10 keV). We chose randomly an N$_H$ from the Gilli et al. 
distribution for each MIPS source. For column densities $\gs 3\times
10^{24}$ cm$^{-2}$ we assumed that the direct emission is completely
blocked by photoelectric absorption and Compton scattering. We used
power law spectra with an energy index equal to 0.8 for the AGN
and 0.9 for the star-forming galaxies. In both cases we assumed a
gaussian dispersion with $\sigma=0.2$ (Ranalli et al. 2003). For the
AGN we also assumed a Compton reflection component with the same
normalization and energy index of the power law component, assuming an
inclination of the reflecting material to the line of sight of 60
degrees, and a scattering component with the same spectral index of
the power law component and normalization 1/100 of that of the power
law component. This is a conservative value which accounts for recent
results on highly obscured AGN with very small low energy scattering
component (Ueda et al. 2007, Comastri et al. 2007).  Fluxes in the
0.3-1.5 keV and 1.5-6 keV band were computed by using the unobscured
2-10 keV luminosities and the assumed spectrum. Finally, count rates
were computed by using the Chandra on axis response.

We run 12 series of simulations (12 different realizations for each of
the sources in each of the 9 cells), varying the fraction of AGN
between 0 and 100\% of the MIPS sources in each cell. This fraction is
then evaluated in each cell by comparing the output of the simulation
with the results of the stacking analysis. Since this analysis was
performed excluding the sources directly detected in X-rays, also
simulations providing more than 10 counts in the 0.3--6 keV band were
excluded from the analysis. Most ($>95\%$) of the simulated AGN with
column densities $\ls10^{24}$ cm$^{-2}$ turned out to have more than
10 counts in the 0.3-6 keV band, and were excluded from the
analysis. This confirms that most unobscured and Compton-thin AGN
would have been detected directly by {\it Chandra}. On the other hand,
simulations with N$_H>10^{24}$ cm$^{-2}$ produced more than 10 counts
$\sim10-20\%$ of the times in cells A and B.  This suggests that some
of our faint Chandra detections might be Compton-thick AGN.

\begin{figure*}[h]
\centering
\begin{tabular}{cc}
\includegraphics[width=8cm, angle=0]{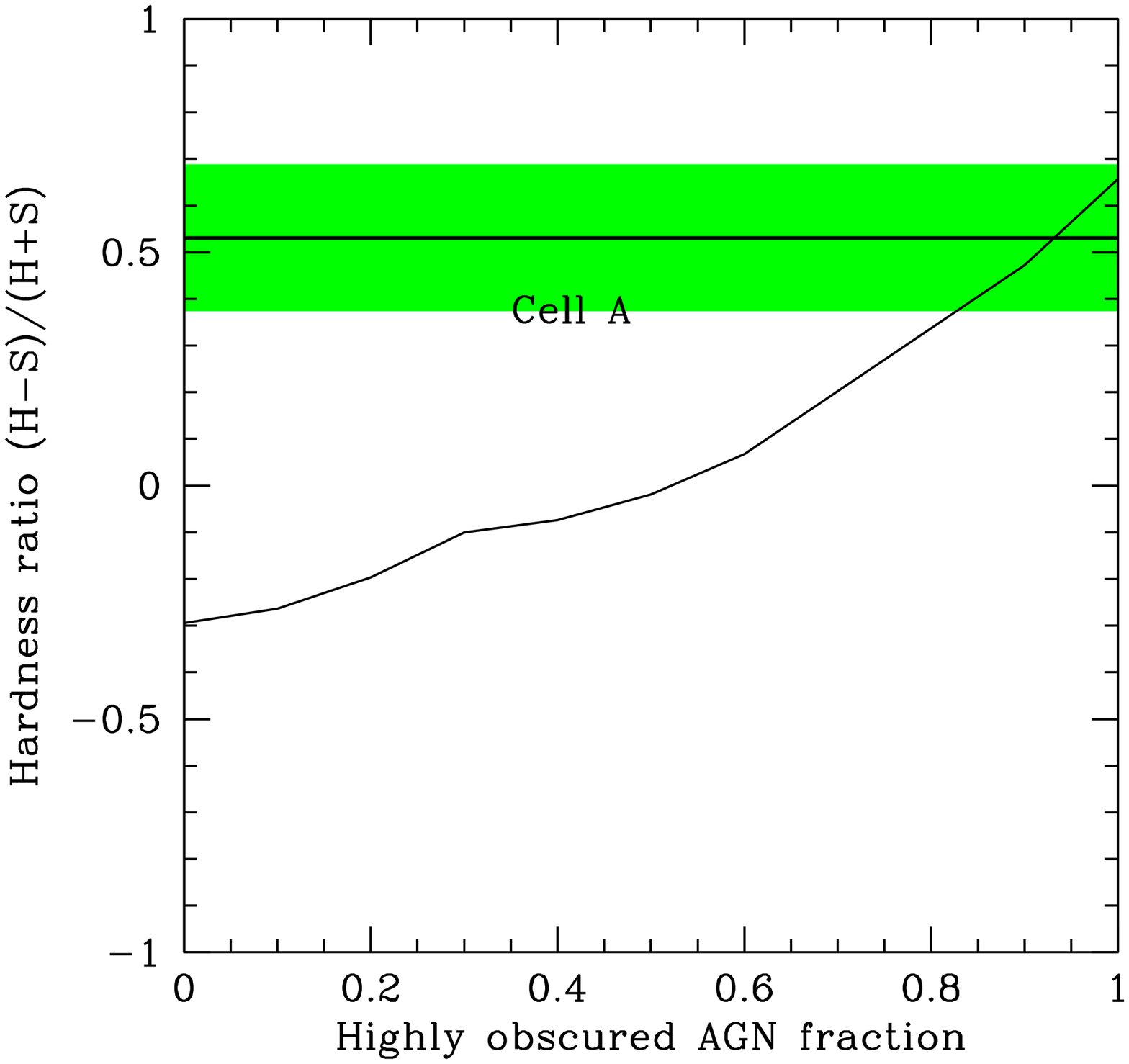}
\includegraphics[width=8cm, angle=0]{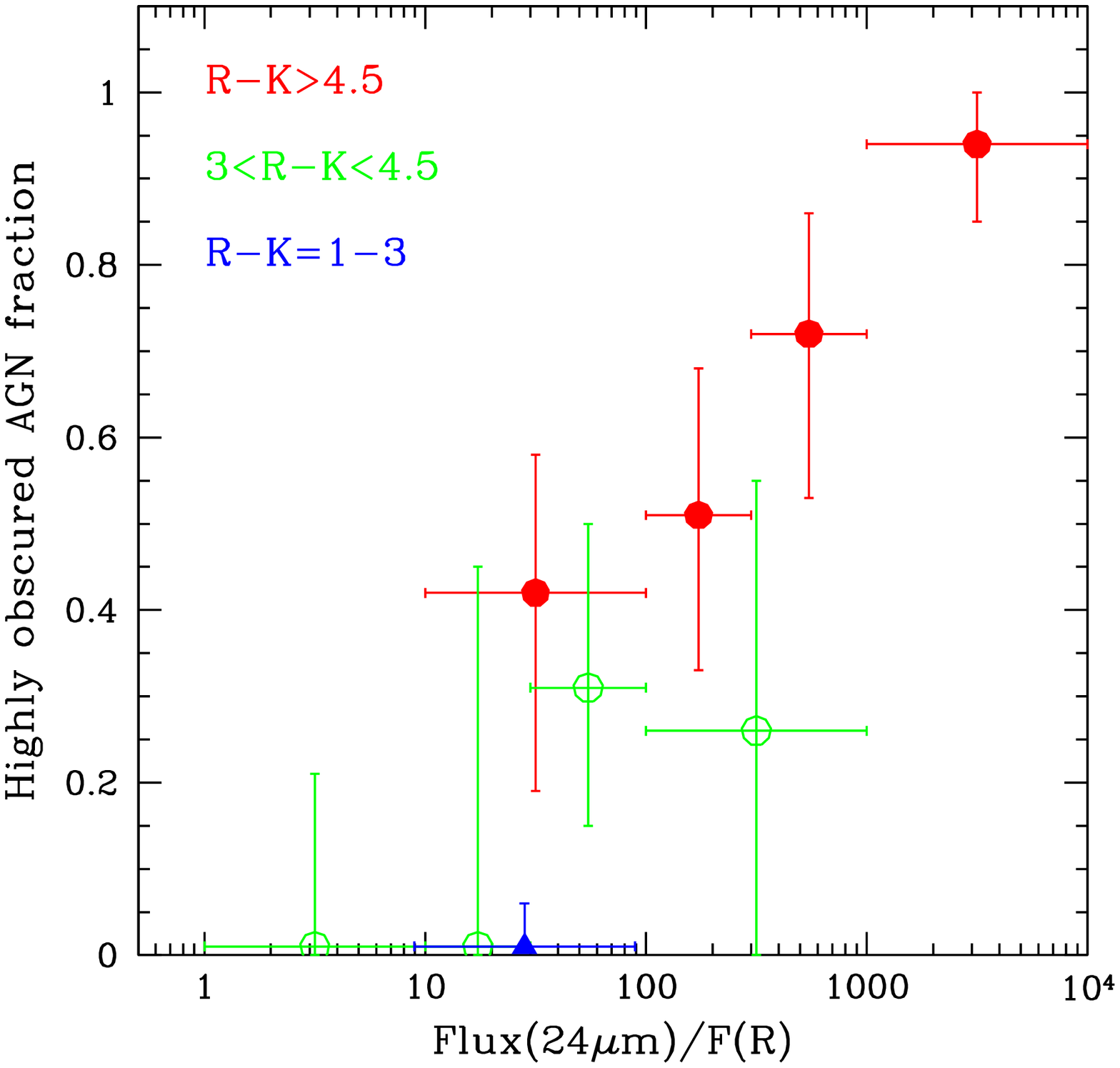}
\end{tabular}
\caption{[Left panel:] the hardness ratio (H-S)/(H+S) as a
function of the fraction of CT AGN in the sample of MIPS sources with
a spectroscopic or photometric redshift in cell A. The solid curve is
the result of Monte Carlo simulations (see text for details); the
thick horizontal lines is the average hardness ratios measured in cell
A.  The colored bands mark the hardness ratio statistical
uncertainties.[Right panel:] The fraction of CT AGN to the total MIPS
source population as a function of F(24$\mu$m)/F(R).  Filled circle =
cells A, B, C , D; open circles = cells E, F, G, H; filled triangle =
cell J.  }
\label{frac}
\end{figure*}

Using these simulations we converted the observed hardness ratios in a
fraction of obscured, most of them CT, AGN to the total MIPS
source population in each of the 9 cells. Fig. \ref{frac} left panel
shows the calibration plot for cell A as an example. Fig. \ref{frac}
right panel shows the derived fraction of obscured AGN to the
total MIPS source population as a function of F(24$\mu$m)/F(R). These
fractions are also given in Table 4.  At the end of the procedure we
verified that the number of sources excluded in each cell is roughly
similar to the number of sources actually detected in the real images.

We studied how the derived fractions vary by changing some of the
assumptions made to build the simulations. The results turned out
qualitatively similar to those reported in Fig. \ref{frac} and Table 4
in all cases. We first studied the impact on the results of a higher
count threshold to exclude entries from the simulations.  Setting the
threshold at 20 counts rather than 10 does not change the fraction of
obscured AGN in cells A and B, while it slightly reduces this fraction
in the other cells.

Assuming a flat N$_H$ distribution with a cut-off at
N$_H=10^{26}$ cm$^{-2}$ instead of the Gilli et al. (2007)
distribution decreases slightly the predicted hardness ratios, thus
increasing the fractions of obscured AGN by 5-10\%. The reason is that
this flat distribution has a fraction of objects in the logN$_H$ bin
25-26 higher than the Gilli et al. distribution.

Assuming a fixed conversion factor between the infrared and X-ray
luminosities equal to that given by eq. 1 at logL(2-10keV)=44 (6.3)
does not change significantly the predicted hardness ratios. Assuming
a fixed factor twice the previous feature, as if additional components
in addition to the nuclear one contribute significantly to the
5.8$\mu$m luminosity, increases slighlty the predicted hardness
ratios, decreasing the fraction of obscured AGN needed to reproduce
the observed ratios by $\sim5\%$.

The main parameter affecting the predicted hardness ratio, in addition
to the fraction of highly obscured AGN, is the energy index assumed
for the star-forming galaxies. Assuming a 20\% steeper (or flatter)
spectral index for these objects increases (or decreases) the
fractions of obscured AGN by $\sim10\%$. The hardest hardness ratios
are observer in cells A and B. These ratios can be produced by a power
law spectrum with energy index -0.5 and 0 respectively reduced at low
energy by Galactic absorption only. Should star-forming galaxies at
z$\sim1.9$ and $\sim1$ (the median redshift of the sources in cells A
and B respectively), exhibit such extremely hard spectra, they will be
able to explain the observed hardness ratios.

\section{SED fittings}

To characterize the infrared SED of the MIPS selected sources we
fitted the SED with a library of empirical templates (Polletta et
al. 2007, Pozzi et al. 2007, F08, Salvato et al. 2008), fixing the
redshift to the spectroscopic redshift or, if this is not available,
to the photometric redshift. To limit the importance of dust
extinction we limited the analysis to the observed bands with
wavelengths $\gs 2\mu$m.  We used the 25 templates of F08 plus
additional templates of hybrid galaxy plus AGN sources by Salvato et
al. (2008).  Table 5 and 6 give the number of sources best fitted by
four broad template categories, in each of the nine cells for sources
with and without a direct X-ray detection. We counted as obscured AGN
sources best fitted by the Salvato et al. (2008) hybrid models with an
AGN contribution larger than 50\%. Sources best fitted by hybrid
templates with a smaller AGN contribution are counted as passive or
star-forming galaxies.  Note that the hybrid models of Salvato et
al. (2008) are normalized around 1$\mu$m, and therefore templates with
an AGN contribution $>$ than 50\% at this wavelength may still have
the bolometric luminosity domininated by star-light.  We see that most
sources in cell A of both samples are best fitted by obscured AGN
templates (81\% and 75\% respectively).

Concerning the sample with a direct X-ray detection, the fraction of
obscured or unobscured AGN to the total MIPS source population
is dominant in all cells but G and D, where the majority of best
fit templates are those of star-forming  galaxies. Unobscured AGN templates
are numerous in cell J, consistently with the spectroscopic
identification of the X-ray sources in this cell (83\% of the
identified sources are type 1 AGN).

Conversely, the number of best fit unobscured AGN templates among the
samples of MIPS sources without a direct X-ray detection is small in
all cells, as expected, since these objects would be more easily
directly detected in X-rays.  The fraction of best fit, obscured
AGN templates to the total is high in cells A and C, while the number
of best fit, star-forming galaxies templates is high in cells D, F and
especially in cell G, where the totality of best fit templates are
those of star-forming galaxies (Table 6).

In conclusion, the result of the SED fitting is qualitatively
consistent with the results obtained from the analysis of the X-ray
properties of the MIPS sources presented in the previous sections and
summarized in Fig. \ref{frac}.

\begin{table*}
\caption{\bf Template fits to the infrared SEDs of MIPS sources with 
a direct X-ray detection}
\begin{tabular}{lccccccccc}
\hline
\hline
Template                     & A  & B  & C  & D  & E  & F  & G  & H & J \\
\hline
%\multicolumn{10}{c}{Passive galaxies}\\         
Passive galaxies$^a$         & 0  & 0  & 0  & 0  & 0  & 1  & 0  & 1 & 0 \\
\hline
%\multicolumn{10}{c}{Star-forming galaxies}\\
Star-forming galaxies$^b$    & 0  & 3  & 2  & 9  & 6  & 15 & 18 & 3 & 1 \\
\hline
%\multicolumn{10}{c}{Galaxies hosting highly obscured AGN and QSOs}\\
Obscured AGN and QSOs$^c$    & 25 & 20 & 21 & 5  & 15 & 33 & 1  & 0 & 19 \\
\hline
%\multicolumn{10}{c}{Unobscured AGN and QSOs}\\
Unobscured AGN and QSOs$^d$  & 6  & 9  & 1  & 0  & 5  & 12 & 0  & 0 & 19 \\
\hline
Total                        & 31 & 33 & 24 & 14 & 26 & 61 & 19 & 4 & 41 \\
\hline
\end{tabular}

$^a$Elliptical + S0 + hybrid passive with AGN contribution $<50\%$;
$^b$Spiral + M82 + Arp220 + N6090 + hybrid with AGN contribution
$<50\%$; $^c$Seyfert 1.8 + Seyfert 2 + red QSOs + I19254 + Mark231 +
A2690\_75 + BPM16274\_69 + IRAS09104+4109 + NGC6240 + hybrid passive
and active with AGN contribution $\geq50\%$; $^d$ Seyfert 1 + QSOs.
\end{table*}

\begin{table*}
\caption{\bf Template fits to the infrared SEDs of MIPS sources without
a direct X-ray detection}
\begin{tabular}{lccccccccc}
\hline
\hline
Template                     & A  & B   & C  & D  & E  & F  & G  & H  & J \\
\hline
%\multicolumn{10}{c}{Passive galaxies}\\         
Passive galaxies$^a$         & 0  & 0   & 0  & 0  & 0  & 1  & 0  & 0  & 0 \\
\hline
%\multicolumn{10}{c}{Star-forming galaxies}\\
Star-forming galaxies$^b$    & 9  & 31  & 16 & 42 & 19 & 65 & 125& 10 & 1 \\
\hline
%\multicolumn{10}{c}{Galaxies hosting highly obscured AGN and QSOs}\\
Obscured AGN and QSOs$^c$    & 30 & 28  & 63 & 8  & 25 & 59 & 0  & 0  & 9 \\
\hline
%\multicolumn{10}{c}{Unobscured AGN and QSOs}\\
Unobscured AGN and QSOs$^d$  & 3  & 3   &  0 & 0  & 2  & 0  & 1  & 0  & 1 \\
\hline
Total                        & 40 & 64  & 81 & 51 & 46 & 125& 125& 10 & 11 \\
\hline
\end{tabular}

$^a$Elliptical + S0 + hybrid passive with AGN contribution $<50\%$;
$^b$Spiral + M82 + Arp220 + N6090 + hybrid with AGN contribution
$<50\%$; $^c$Seyfert 1.8 + Seyfert 2 + red QSOs + I19254 + Mark231 +
A2690\_75 + BPM16274\_69 + IRAS09104+4109 + NGC6240 + hybrid passive
and active with AGN contribution $\geq50\%$; $^d$ Seyfert 1 + QSOs.
\end{table*}

\section{The AGN fraction}

\begin{figure*}[t!]
\centering
\begin{tabular}{cc}
\includegraphics[width=8cm, angle=0]{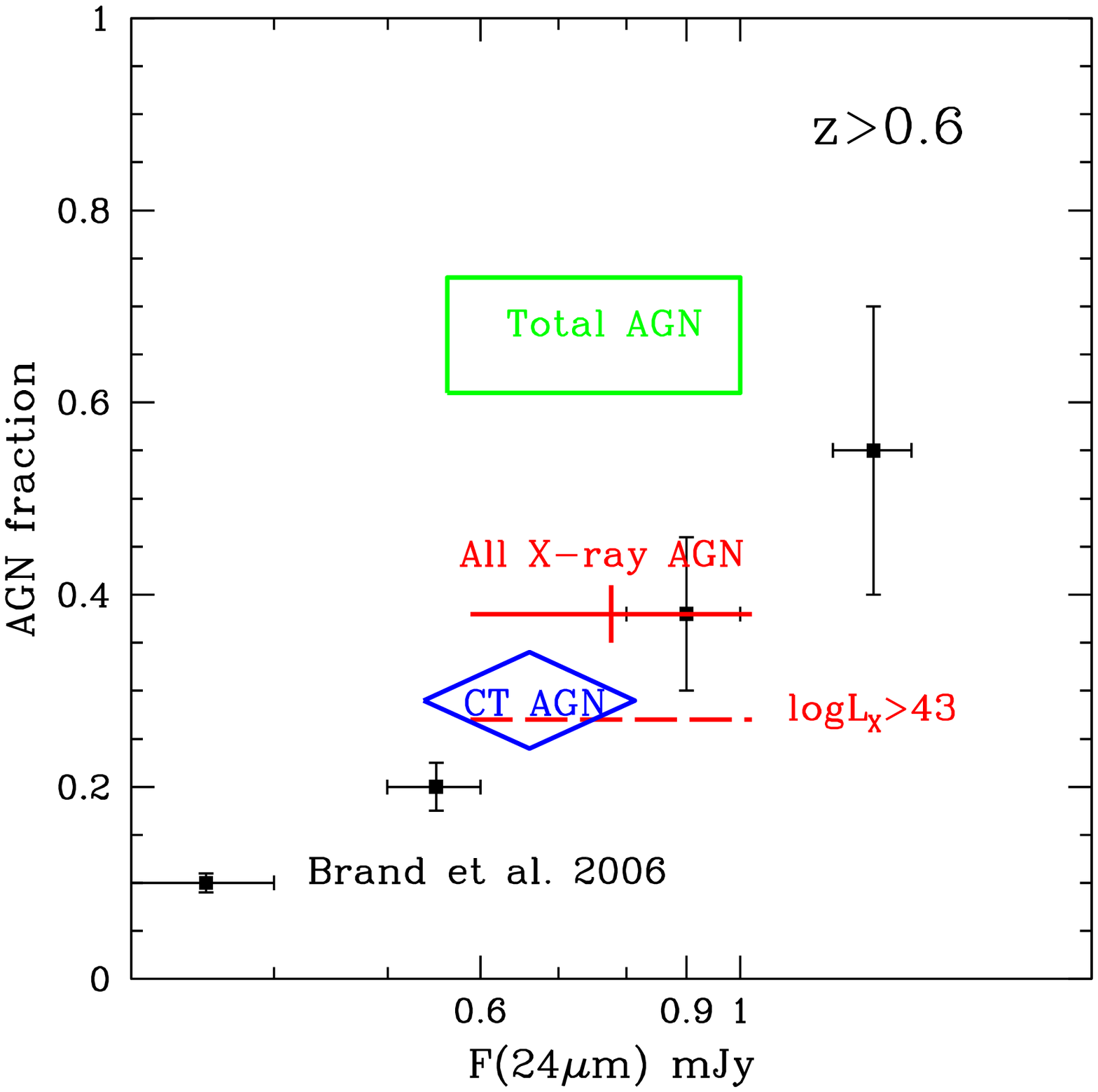}
\includegraphics[width=8cm, angle=0]{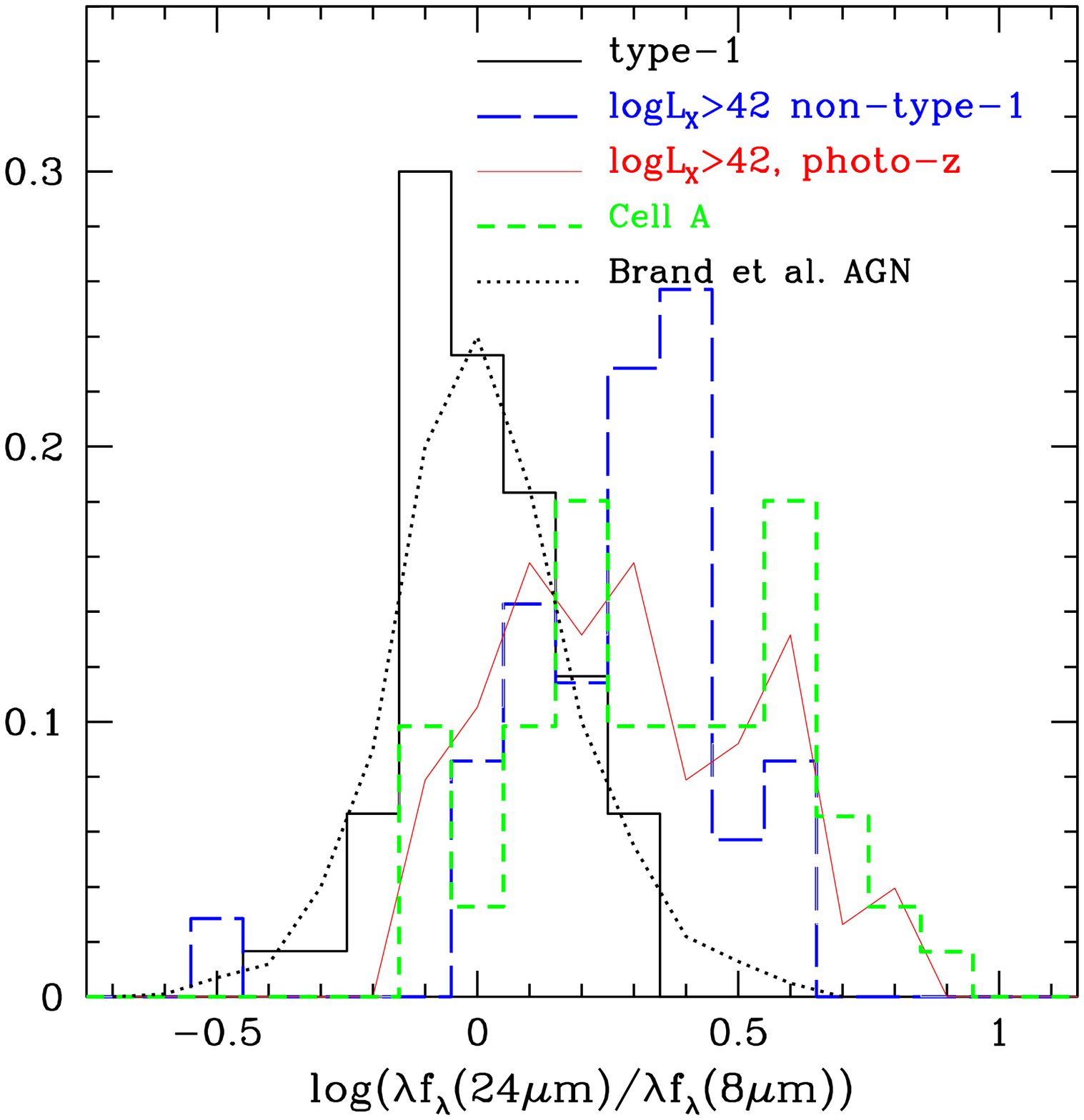}
\end{tabular}
\caption{[Left panel:] The AGN fraction as a function of the 24$\mu$m flux
for z$>0.6$ sources.  Red cross and dashed line = AGN with a direct
X-ray detection; blue diamond = CT AGN without a direct X-ray
detection; green box = total AGN fraction. The black points are the
AGN fraction of Brand et al. (2006) [Right panel:] The distribution of the
F(24$\mu$m)/F(8$\mu$m) flux ratio for several COSMOS MIPS samples at
z$>0.6$: black, solid histogram = type 1 AGN with a direct X-ray
detection; blue, long-dashed histogram = type 2 AGN with a direct
X-ray detection; red thin solid line = sources with photometric
redshift and an X-ray luminosity $>10^{42}$ ergs s$^{-1}$; green,
short-dashed histogram = sources in Cell A without a direct X-ray
detection. The Brand et al. (2006) X-Bootes AGN distribution is showed
as a comparison (black, dotted line). }
\label{fractot}
\end{figure*}

We can now evaluate the total AGN fraction (including unobscured,
moderately obscured and highly obscured AGN) in a 24$\mu$m source
sample. About $75\%$ of the MIPS sources with a direct Chandra
detection have an X-ray (2-10 keV) luminosity $>10^{42}$ ergs $^{-1}$,
and are therefore likely to host an AGN. This already makes $\sim23\%$
of the full COSMOS MIPS sample.  Taken at face values, the infrared
selected, CT AGN fractions given in Table 4 would imply that
$36\pm11\%$ of the COSMOS MIPS sources without a direct Chandra
detection host an AGN ($\sim26\%$ of the full MIPS sample) a fraction
similar to that of the X-ray detected AGN.  The total fraction of AGN
in the full MIPS sample, obtained by adding the two previous
fractions, is $49\pm10\%$.  At the typical 24$\mu$m fluxes of the
COSMOS bright MIPS sample (flux limit $\sim550\mu$Jy, median flux
$\sim750\mu$Jy) this is quite a large fraction, compared with previous
studies (see Brand et al. 2006 and references therein). A more
accurate comparison with the study of Brand et al. (2006) can be made
by considering the sources at z$>0.6$ only (Brand et al. use this
redshift cut to avoid contamination from low-z star-forming
galaxies). The left panel of Fig. \ref{fractot} shows the AGN
fractions in the COSMOS MIPS bright sample at z$>0.6$ for both sources
with a direct X-ray detection and CT AGN, selected using
infrared/optical colors and the {\it Chandra} stacking analysis
presented in Sections 3.2 and 3.3.  The fraction of MIPS AGN with a
direct X-ray detection is already higher than the Brand et al. (2006)
estimates at the same 24$\mu$m fluxes. The fraction of AGN with X-ray
luminosity higher than $10^{43}$ ergs s$^{-1}$ is consistent with the
Brand et al. estimates. The fraction of X-ray AGN is also of the same
order of magnitude of that of CT AGN.  Adding the fractions of X-ray
selected and CT AGN results in a total fraction of AGN in the COSMOS
bright MIPS sample at z$>0.6$ of $0.67\pm0.06$, a factor of 2 higher
than the Brand et al. (2006) estimates. The reason for this apparent
inconsistency is that while Brand et al. (2006) assume that the AGN
dominates the bolometric luminosity, in many of the sources of our
MIPS sample the AGN the bolometric luminosity is dominated by the host
galaxy.  This can be also appreciated by considering the right panel
of Fig. \ref{fractot}. It shows the distribution of the
F(24$\mu$m)/F(8$\mu$m) flux ratio for several COSMOS MIPS samples at
z$>0.6$. X-ray selected, type 1 AGN peak around
log(F(24$\mu$m)/F(8$\mu$m))=0, and their distribution is consistent
with that of X-ray detected AGN in the Bootes field. On the other
hand, the distribution of X-ray selected AGN with an optical spectrum
not showing broad lines (non type 1 AGN), and that of X-ray AGN with
only a photometric redshift, are significantly shifted toward higher
log(F(24$\mu$m)/F(8$\mu$m) values. Note as the latter distribution is
similar to the distribution of sources in Cell A without a direct
X-ray detection. Brand et al. (2006) associate to the AGN population
the peak of the MIPS source distribution around
log(F(24$\mu$m)/F(8$\mu$m)=0 and therefore miss part of the X-ray
selected AGN without broad lines in their optical spectra or without
optical spectra, and CT AGN without a direct X-ray detection with a
higher log(F(24$\mu$m)/F(8$\mu$m) ratio.  Both populations could be
recovered in the COSMOS field thanks to the much deeper X-ray coverage
(a {\it Chandra} exposure time 20-40 times longer than that on the
BOOTES field), which allowed the direct detection of Seyfert 2 like
galaxies up to z$\sim1$ and a detailed stacking analysis of the MIPS
sources without a direct X-ray detection.

\section{The density of CT AGN}

In the previous Sections we estimated the fraction of obscured AGN
to the total MIPS source population, not detected directly in
X-rays, but visible in the stacked {\it Chandra} images, in each of
the 9 cells defined in the F(24$\mu$m)/F(R) -- R-K diagram. Our
simulations showed that most of these AGN are likely to be CT and
therefore we will call them CT AGN (or CT QSOs when referring to high
luminosity sources) for the sake of simplicity. We can now compute the
volume density of the MIPS selected sources, correcting for this
fraction, to obtain the density of CT AGN in different redshift and
luminosity bins. We use for this calculation the standard 1/V$_{max}$
method (Schmidt 1968, Lilly et al. 1995, Cowie et al. 2003). While it
is well known that this method is not free from biases (the main one
is that it does not account for evolution within each L and z bin), it
is robust enough to derive general trends (see e.g. Cowie et
al. 2003).

\begin{figure*}[t!]
\centering
\begin{tabular}{cc}
\includegraphics[width=8cm, angle=0]{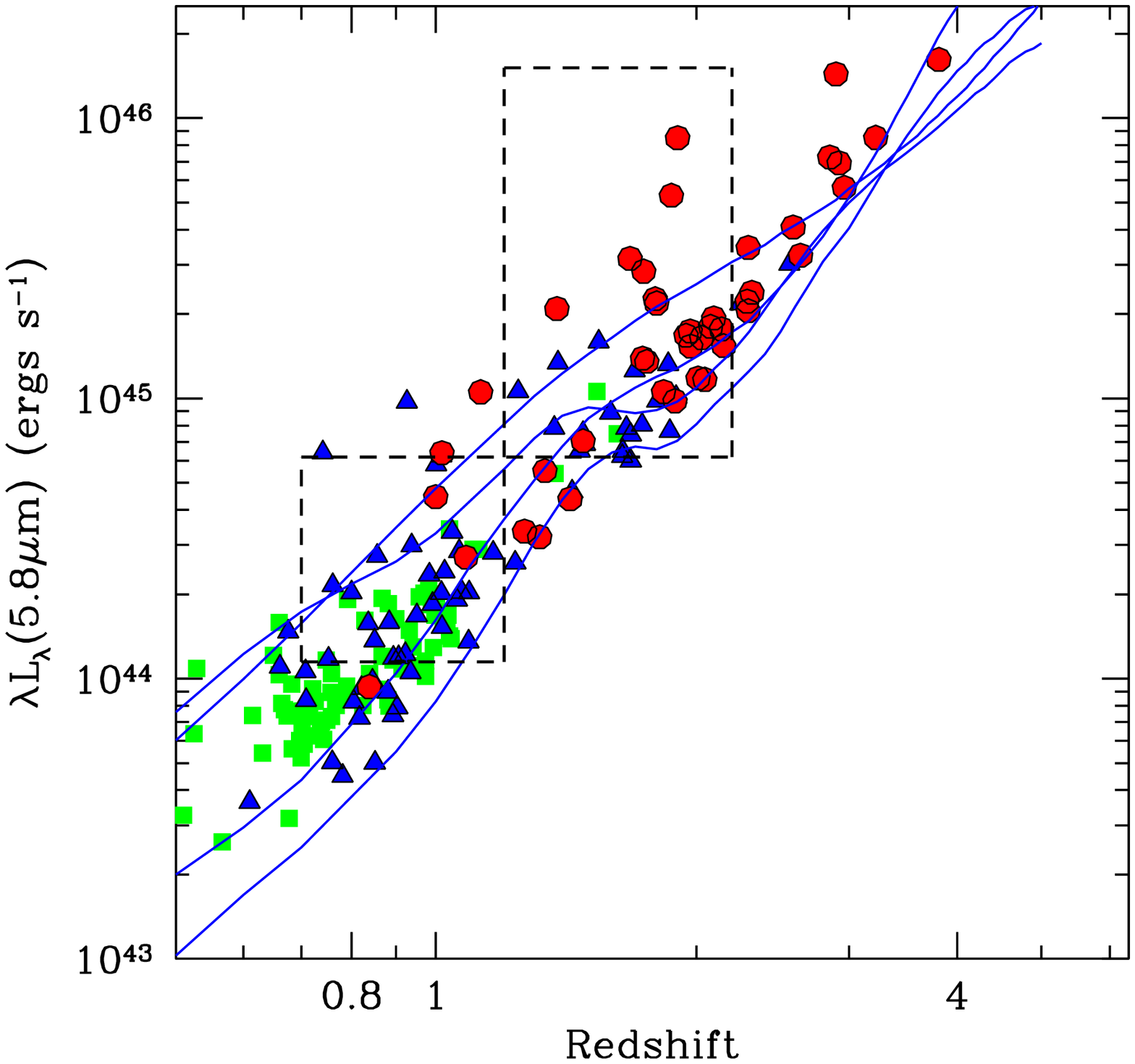}
\includegraphics[width=8cm, angle=0]{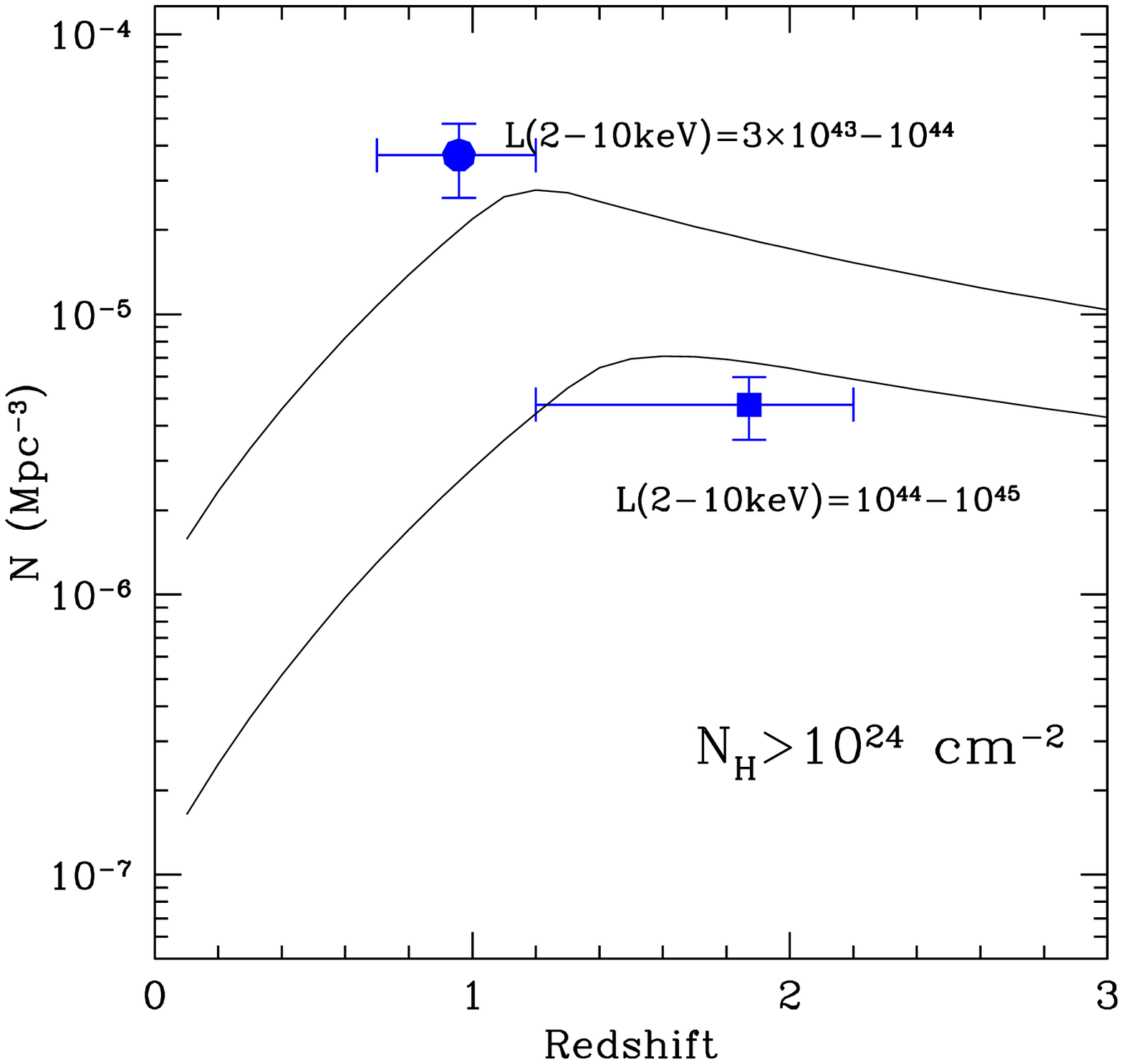}
\end{tabular}
\caption{[Left panel:] the redshift-luminosity plane of the MIPS
sources in cell A (red circles), cell B (blue triangles) and cell
C (green squares).  Two redshift-luminosity bins chosen for the
computation of the CT AGN volume density are marked as
dashed boxes. The solid curves show the loci of four SEDs of highly
obscured AGN for the 24$\mu$m flux limit of 550$\mu$Jy. [Right panel:]
volume density of COSMOS MIPS selected CT AGN. Solid
curves are the expectation of the Gilli, Comastri \& Hasinger (2007)
model in two luminosity bins.}
\label{evol}
\end{figure*}

Fig. \ref{evol} left panel shows the infrared luminosity - redshift
plane for the MIPS sources without a direct X-ray detection in cells
A, B and C. The loci of four SEDs of highly obscured AGN computed at
the 24$\mu$m flux limit of 550$\mu$Jy are also showed. Two
redshift-luminosity bins chosen for the computation of the CT AGN
volume density are also marked in figure. The two bins have been
chosen according to the following three criteria:

\begin{enumerate}
\item
The bins must lie above the lowest limit for an obscured AGN SED
corresponding to our 24$\mu$m flux limit.  We can therefore considered
the source samples in these bins relatively little affected by complex
selection effects.

\item
The redshift and luminosity ranges should not be too large, to avoid
strong biases in the calculation of the volume densities, see above.

\item
The bins should be cut to maximize the number of sources included in
the analysis at the relevant redshifts, to keep statistical errors
small.
\end{enumerate}

Fig. \ref{evol}, left panel, shows that a reasonable compromise is to
limit the analysis to the sources in the redshift-luminosity bins given 
in Table 7.  This table gives for each of these bins the number of MIPS
sources without a direct X-ray detection in some of the nine
F(24$\mu$m)/F(R) and R-K cells (the total number of sources in each
cell is given in brackets in the last column of Table 7).

\begin{table}
\caption{\bf Number of sources in luminosity-redshift bins}
\begin{tabular}{lccc}
\hline
\hline
Redshift & log($\lambda L_\lambda(5.8\mu$m)) & Cell & \# of sources\\
         &                                   &      &              \\
\hline
$1.2-2.2$  & 44.79-46.18    &  A    &  21 (40)\\
$1.2-2.2$  & 44.79-46.18    &  B    &  15 (64)\\
$1.2-2.2$  & 44.79-46.18    &  C    &  2 (81)\\
$1.2-2.2$  & 44.79-46.18    &  D    &  0 (51)\\
$1.2-2.2$  & 44.79-46.18    &  E+F+G+H &  3 (306)\\
$0.7-1.2$  & 44.06-44.79    &  A    &  2  (40)\\ 
$0.7-1.2$  & 44.06-44.79    &  B    &  25 (64)\\ 
$0.7-1.2$  & 44.06-44.79    &  C    &  29 (81)\\ 
$0.7-1.2$  & 44.06-44.79    &  D    &  2 (51)\\ 
$0.7-1.2$  & 44.06-44.79    &  E+F+G+H &  11 (306)\\ 
\hline
\end{tabular}

\end{table}

Because of the strong correlation of F(24$\mu$m)/F(R) (and therefore
R-K) with log($\lambda L_\lambda(5.8\mu$m)) the largest numbers of
sources with large infrared luminosity are found in cells A and B,
which are also the cells with the highest fraction of CT AGN. On the
other hand, the largest number of sources with intermediate luminosity
at z=0.7--1.2 are found in cell C. The contribution of sources in
cells E,F,G and H is small in both luminosity-redshift bins.  The
volume densities of CT AGN and CT QSOs are computed using the
sources in Table 7, a $24\mu$m flux limit of 550$\mu$Jy and correcting
the resulting density of MIPS sources for the fraction of CT AGN in
Table 4. The densities are reported in Table 8 and in Fig. \ref{evol}
(right panel). The two L($5.8\mu$m) luminosity bins in Table 8
corresponds to unobscured 2-10 keV luminosities of
$3\times10^{43}-10^{44}$ and $10^{44}-10^{45}$ ergs s$^{-1}$ using the
L($5.8\mu$m)-L(2-10 keV) luminosity conversion of eq. 1.

\begin{table*}
\caption{\bf AGN volume densities}
\begin{tabular}{lccccc}
\hline
\hline
Redshift & log($\lambda L_\lambda(5.8\mu$m)) & logL(2-10keV) & CT AGN  &  
Tot. X-ray AGN$^a$ & Unobscured X-ray AGN$^a$\\
         &          ergs s$^{-1}$            & ergs s$^{-1}$ &  Mpc$^{-3}$ & 
Mpc$^{-3}$ & Mpc$^{-3}$ \\
\hline
$1.2-2.2$  & 44.79-46.18    & 44-45        & $(4.8\pm1.1)\times10^{-6}$  &
$1.1\times10^{-5}$ & $5.4\times10^{-6}$\\
$0.7-1.2$  & 44.06-44.79    & 43.477-44    & $(3.7\pm1.1) \times10^{-5}$ &
$5.4\times10^{-5}$ & $3.0\times10^{-5}$\\
\hline
\end{tabular}

$^a$ evaluated using the La Franca et al. 2005 luminosity function 
parameterization.
\end{table*}

\begin{table*}
\caption{\bf Previous CT AGN volume densities determinations}
\begin{tabular}{lccll}
\hline
\hline
Paper & Field    & Redshift & Luminosity & CT AGN density \\
      &          &          &  ergs s$^{-1}$ & \\
\hline
Daddi et al. 2007 & CDFS    & 1.4--2.5 & L(2-10keV)=$(1-4)\times10^{43}$ &
$\approx2.6\times10^{-4}$ Mpc$^{-3}$ \\
Fiore et al. 2008 & CDFS    & 1.2--2.6 & logL(2-10keV)$\gs43$  &
$\sim 100\%$ X-ray selected AGN\\
Donley et al. 2008 & CDFS   &   --     &  --                  &
54--94\%$^a$ X-ray selected AGN  \\
Alexander et al. 2008 & CDFN & 2--2.5  & logL(2-10keV)=44--45 &
$0.7-2.5\times 10^{-5}$ Mpc$^{-3}$ \\
Martinez-Sansigre et al. 2007 & SWIRE SXDS & 1.7--4 & logL$_{bol}\gs47$ & 
$\gs$ unobscured QSOs \\ 
Polletta et al. 2008 & SWIRE, NDWFS, FLS & 1.3--3 &  
L($6\mu$m)$\gs4\times10^{45}$ & 37-65\%$^b$ total AGN population\\ 
Della Ceca et al. 2008 & XMM HBS & 0 & logL(2-10keV)=43 &
$(0.8-2.8)\times10^{-5}$  Mpc$^{-3}$\\
Della Ceca et al. 2008 & XMM HBS & 0 & logL(2-10keV)=44 &
$(1-5)\times10^{-7}$ Mpc$^{-3}$ \\
Della Ceca et al. 2008 & XMM HBS & 0 & logL(2-10keV)=45 &
$(1-25)\times10^{-10}$ Mpc$^{-3}$  \\
\hline
\end{tabular}

$^a$ The lower limit refers to {\it Spitzer} selection only, the upper
limit includes the contribution of AGN selected because their high
radio to infrared flux ratio; $^b$ The lower limit refers to QSO
obscured by a compact torus, the upper limit to the global fraction of
obscured QSO to the total QSO population in that redshift-luminosity
bin.
\end{table*}

\section{Discussion}

We analyzed the X-ray properties of a sample of MIPS 24$\mu$m sources
with a signal to noise ratio $>4$ and a 24$\mu$m flux limit of
$\sim550 \mu$Jy detected in an area of the COSMOS field with deep {\it
Chandra} coverage.  232 of the 919 MIPS sources have a robust {\it
Chandra} detection (Elvis et al. 2008). Additional 47 sources have
more than 10 background-subtracted {\it Chandra} counts within 5
arcsec of the position of the MIPS source, but they are not present in
the C-COSMOS and XMM-COSMOS catalogs. These may well be faint X-ray
sources just below the adopted detection threshold. The fraction of
X-ray detections for the COSMOS bright MIPS sample is therefore
between 25\% and 30\%.  $\sim75\%$ of these sources have an X-ray
(2-10 keV) luminosity $>10^{42}$ ergs $^{-1}$, and are therefore
likely to host an AGN. For the sources not directly detected in the
{\it Chandra} images we computed stacked count rates in the 0.3-1.5
keV and 1.5-6 keV bands.  We found a strong correlation between the
hardness ratio of the stacked count rates and the F($24\mu$m)/F(R)
flux ratio (and the R-K color). We use the stacked count rates and
hardness ratios, together with detailed Monte Carlo simulations, to
estimate the fraction of Compton-thick AGN in nine cells defined in
the F(24$\mu$m)/F(R) -- R-K diagram. The simulations were performed
assuming that the sources without an X-ray detection are either
obscured AGN or star-forming galaxies. The results, reported in Table
4 and Fig. \ref{frac}, suggest that a large fraction of `IR excess'
sources should host an obscured but active nucleus. In particular, the
hard hardness ratios measured in cells A and B, those with the highest
F($24\mu$m)/F(R) flux ratio (Fig. \ref{hrmiroc}), would imply a
fraction of Compton-thick AGN as high as 0.93--0.73, respectively.
These results were obtained by using reasonable assumptions for the
AGN and star-forming galaxies X-ray spectra and their normalization to
the 5.8$\mu$m luminosity. In particular, we adopted for the
star-forming galaxies a power law X-ray spectrum with an energy
spectral index 0.9, similar to that found in star-forming galaxies
accessible to detailed X-ray spectroscopy. To explain the observed
hardness ratios without resorting to a large population of CT AGN
would require extremely flat or even inverted energy spectral indices
for most star-forming galaxies at z=1-2, a spectrum never observed so
far in this class of objects at smaller cosmological distances.  This
would imply a rather extreme cosmological evolution of the X-ray
spectrum of star-forming galaxies, that, while cannot be fully
excluded, appears nevertheless unlikely.  Furthermore, the
interpretation of the observed hardness ratio in terms of a high
fraction of CT AGN is fully consistent with the results of template
fitting to the observed MIPS source's SEDs.

Recently Donley et al. (2008) questioned the `IR excess' technique to
select highly obscured AGN, claiming that CDFS 'IR excess' samples may
be more contaminated by moderately obscured AGN and star-forming
galaxies than estimated by Daddi et al. (2007) and F08. For the
purposes of this paper we limit ourselves to remark that this can
hardly be the case for luminous QSOs in cell A. In fact, an unobscured
QSO with L(2-10keV)$=10^{44}$ ergs s$^{-1}$ at z=2.2 (the upper limit
of the redshift bin considered in Section 5), would produce between 40
and 80 counts in the C-COSMOS images for the two typical exposure
times in Table 2, and it would be easily detected.  A QSO with the
same redshift and luminosity but with a column density of
$5\times10^{23}$ \cm2 would still produce between 10 and 20 counts,
thus having a high probability of being detected. Should this source
be obscured by a column density as high as $10^{24}$ \cm2 it would
still produce between 6 and 12 counts (direct emission only, without
considering the likely contribution from a scattering component), with
a non negligible probability of being directly detected. We conclude
that unobscured or moderately obscured QSOs cannot be present in the
sample of MIPS source without a direct X-ray detection at z=1.2-2.2
and logL(2-10keV)=44--45. This sample may well contain star-forming
galaxies. However, if this component is the dominant one, then it
would be difficult to explain the high hardness ratio measured for
these sources (see above). Similar arguments apply to the sources in
the redshift bin 0.7--1.2 and luminosity bin logL(2-10keV)=43.477-44,
which are mainly located in cells B and C (see Table 7).  We conclude
that the 'IR excess' selection appears quite robust, at least
regarding AGN with intermediate to high luminosity at z=0.7--2.2.

\subsection{The cosmic evolution of obscured AGN}

The bright flux limit of the COSMOS bright MIPS sample allows a
rather limited coverage of the luminosity-redshift plane, see
Fig. \ref{evol}, left panel.  Nevertheless, we could select two
redshift-luminosity bins in which the 24$\mu$m source samples can be
considered reasonably complete. This allows us to search for cosmic
evolution of obscured AGN, not directly detected in X-rays.

Table 8 gives the volume densities of CT AGN in the two
redshift-luminosity bins.  They were calculated by correcting the
volume density of the MIPS 24$\mu$m source for the fraction of
obscured AGN not directly detected in X-rays given in Table 4.  It
should be noted that these densities do not account for CT objects
directly detected in X-rays. The identification of a CT spectrum in a
faint X-ray source is not a straight-forward task. Previous studies
suggest that the fraction of CT AGN in X-ray samples is small, of the
order of a few \% (Tozzi et al. 2006, Mainieri et al. 2007, Perola et
al. 2004), so we do not try to correct our density of CT QSOs for
this fraction at this stage.

\subsubsection{The fraction of obscured AGN as a function of their 
luminosity}

Table 8 also gives the volume densities of X-ray selected AGN in the
same redshift bins, computed using the parameterization of the 2-10 keV
luminosity function in La Franca et al. (2005). We find that the
density of infrared selected, CT QSOs at z-1.2--2.2 is 44\% of that of
all X-ray selected AGN in the same redshift-luminosity bin ($\sim90\%$
of that of both unobscured and moderately obscured QSOs). 
Conversely, at z=0.7--1.2 and L(2-10keV)$=3\times
10^{43}-10^{44}$ the density of infrared selected, CT AGN is
$\sim67\%$ of that of X-ray selected AGN, and 120\%, 150\% that of
unobscured and moderately obscured AGN respectively. This comparison
suggests that the fraction of obscured AGN to the total AGN population
decreases with the luminosity not only when considering moderately
obscured, X-ray selected AGN (Ueda et al. 2003, La Franca et
al. 2005), but also including infrared selected, CT AGN.

It is also instructive to compare our estimates to the expectations of
AGN synthesis models for the CXB.  The expectations of the Gilli et
al. (2007) model in the two luminosity bins used in our analysis are
plotted in Fig \ref{evol}. We see that the Gilli et al.  model
predicts a density of CT QSOs (logL(2-10keV)=44-45) slightly higher
than our estimate in the redshift bin 1.2--2.2, but consistent with it
at the 90\% confidence level. Conversely, the model predicts a density
of logL(2-10keV)=43.477--44 AGN at z=0.7--1.2 a factor of $\sim2$
lower than our estimates.  However, this difference is significant at
1.5$\sigma$ level.  The Gilli et al. (2007) model predicts that the
fraction of obscured AGN (logN$_H>22$ \cm2, including CT sources) to
the total AGN population decreases with luminosity from $\sim75\%$ at
Seyfert like luminosities to 45\% at QSO luminosities. In summary, our
determinations of the CT AGN densities in two luminosity bins agree
reasonably well with the Gilli et al. (2007) prediction
(Fig. \ref{evol}). This means again that our findings support the
idea that CT AGN follow a luminosity dependence similar to Compton
thin AGN, becoming relatively rarer at high luminosities.

\subsubsection{Comparison with previous studies.}

We compare our estimates of the CT AGN volume density with
previous determinations (see Table 9 for a summary).

Several recent papers have been focusing on the search for highly
obscured AGN in the CDFS.  Using an approach similar to ours, i.e
comparison of the ratio of the counts in stacked images in two energy
bands with Monte Carlo simulations including both obscured AGN and
star-forming galaxies, F08 estimated in the CDFS a density of CT AGN
with logL(2-10keV)$>43$ and at z=1.2--2.6 similar to that of X-ray
selected AGN.  This cannot be directly compared with the densities
estimated in this paper, because in the same redshift bin we can
select only luminous QSOs. However, we note that the F08 estimate is
similar to what we find in the C-COSMOS field in the lower redshift
bin 0.7--1.2.  Daddi et al. (2007), using a somewhat different
approach on the same CDFS dataset, estimate that the density of CT AGN
is $\sim2.6\times10^{-4}$ Mpc$^{-3}$ at z=1.4--2.5, and infer that
their 2-10 keV luminosities are in the range
$10^{43}-4\times10^{43}$. The Daddi et al. density is significantly
higher than our estimates, and it is $\sim6$ times higher than the
expectation of the Gilli et al. (2007) model.  Donley et al. (2008)
recently estimated a conservative lower limit to the {\it
Spitzer}-selected AGN in the CDFS, not directly detected in
X-rays. They conclude that the number of AGN with 24$\mu$m flux higher
than 80$\mu$Jy is 54-77\% larger than for purely X-ray selected AGN.
The fraction increases to 71-94\% including AGN selected with a
high radio to infrared flux ratio. The combined analysis of the COSMOS
and CDFS field is clearly needed for a better coverage of the
redshift-luminosity plane, and thus to measure the CT AGN luminosity
function in several luminosity and redshift bins.  This combined
analysis will be presented in a forthcoming paper.

Alexander et al. (2008) used again {\it Chandra} deep fields, but
limited their analysis to CT QSOs confirmed through infrared and
optical spectroscopy, in addition to X-ray imaging.  Using a small
sample of four CT QSOs they estimate a density $0.7-2.5\times 10^{-5}$
Mpc$^{-3}$ at z=2--2.5. This is formally higher than our estimate at a
similar redshift, but still statistically consistent with it, within
their rather large error bars.

Martinez-Sansigre et al. (2005, 2007) and Polletta et al. (2008)
looked at highly obscured QSOs of extreme luminosity in the large area
SWIRE, NDWFS and FLS surveys (bolometric luminosity $\gs10^{47}$ ergs
s$^{-1}$). The former authors concluded that their CT QSOs are at
least as numerous as unobscured QSOs, a result similar to what we find
at slightly lower luminosities in the C-COSMOS field.  On the other
hand, Polletta et al. (2008) conclude that 37-40\% of the QSOs with
L($6\mu$m)$\gs4\times10^{45}$ ergs s$^{-1}$ and z=1.3-3 are obscured
by a compact torus, while 23-25\% are obscured by matter distributed
on larger scale. Polletta et al. (2008) do not distinguish between CT
and Compton-thin absorbers, however it is reasonable to assume than
most objects obscured by a compact torus are CT, and therefore that
their fraction is a lower limit to the real CT fraction.

Finally, Della Ceca et al. (2008) estimated the density of CT AGN at
three luminosities, by comparing the luminosity function of optically
selected, narrow line AGN in the SDSS (Simpson 2005), which must
include both CT and Compton-thin AGN, to the luminosity function of
X-ray selected Compto-thin AGN, rescaled at z=0 using their best fit
evolutionary model.  The Della Ceca et al. (2008) densities are listed
in Table 9. They agree quite well with the densities estimated in this
paper, once they are de-evolved (Della Ceca et al.  2008).

\subsection{Obscured AGN: catching feedback in action}

Our findings indicate that luminous CT absorbers follow the same
fundamental correlation with the luminosity found for Compton-thin
absorbers. This correlation has been interpreted in the past either in
terms of a luminosity dependence of the obscuring ``torus''
sublimation radius (e.g. Lawrence 1991), or due to the bending of the
interstellar gas due to the BH gravitational field (the higher the BH
mass the larger the bending, and smaller the fraction of sight lines
intercepting the gas, Lamastra et al. 2006). Recently, Menci et al.
(2008) introduced a new scenario, in which the absorption properties
of an AGN depends both on the orientation to the line of sight and on
the time needed to sweep the central regions of galaxy disks. In this
scenario the correlation between the fraction of obscured AGN and the
luminosity is mainly due to a different timescale over which nuclear
feedback is at work. If this is the case, then one may expect that
X-ray selected, moderately obscured QSOs are caught at a later stage
of feedback activity than highly obscured, CT QSOs. Of course this
must be intended on average, because the distribution of the obscuring
gas around the nucleus would not be spherically symmetric, and
therefore obscuration will depend also on the orientation to the line
of sight. As a consequence, some CT QSO seen just along the plane of
the obscuring material would be in a similar evolutionary stage as a
moderately obscured QSO.  A prediction of this evolutionary scenario
is that the host galaxies of high luminosity, CT QSOs should be, on
average, more star-forming than the host galaxies of unobscured or
moderately obscured QSOs of similar luminosity.  Interestingly,
Stevens et al.  (2005) and Page et al. (2004) find that X-ray obscured
QSOs have much higher sub-millimeter detection rates than X-ray
unobscured QSOs. Alexander et al. (2005) found that most radio
identified sub-mm galaxies host X-ray and optically obscured AGN, but
that their bolometric luminosity is dominated by star-formation.
Martinez-Sansigre et al. (2005, 2008) found little or no
Lyman-$\alpha$ emission in a sample of z$>1.7$ obscured QSOs,
suggesting large scale (kpc) dust distribution. Sajina et al. (2007)
and Martinez-Sansigre et al. (2008) report Spitzer IRS spectra
dominated by AGN continuum but showing PAHs features in emission in
samples of ULIRGs and radio selected obscured QSOs at
z$\sim2$. Finally, Lacy et al. (2007) find evidence for dust-obscured
star formation in the IRS spectra of type-2 QSOs.  All these findings
are in general agreement with the evolutionary picture.

\subsection {Tests of the AGN/host-galaxy evolutionary scenario} 

We were able to select in the C-COSMOS field both unobscured and
moderately obscured AGN using direct X-ray detection, and highly
obscured, CT AGN selected in the infrared but confirmed through a
detailed X-ray stacking analysis. This make our sample ideal to probe
the above evolutionary scenario.

Star-formation can be revealed by emission at radio wavelengths
(Condon 1992).  Radio fluxes, down to a $5\sigma$ flux limit of $\sim
65\mu$Jy are available for 43\% of the MIPS selected sources. The same
fraction of radio fluxes is available for MIPS sources with a direct
X-ray detection. Following Schinnerer et al. (2007), the 1.4 GHz
luminosity has been computed assuming a spectral energy index of
0.8. Among the 38 sources in cells A and B and in the
redshift-luminosity bin z=1.2--2.2 logL(5.8$\mu$m)=44.79-46.18 (see
Table 7) 18 have a radio detection. Their median logarithmic radio
luminosity ($<$log$\lambda L_\lambda(1.4GHz)$) is 40.53 ergs s$^{-1}$
with interquartile range 0.18. Putting the radio luminosity of the
sources without a radio detection to the limit computed at their
redshift reduces the median logarithmic radio luminosity to 40.29 ergs
s$^{-1}$ with interquartile 0.18. If the radio luminosity is due to
star-formation, it would imply star-formation rates of the order of
300 M$_{\odot}$ yr$^{-1}$ (e.g. using the correlation given by Condon
1992).

The median logarithmic ratio between the 5.8$\mu$m luminosity and the
1.4GHz luminosity of the 38 sources in cells A and B and in the
redshift-luminosity bin z=1.2--2.2 logL(5.8$\mu$m)=44.79-46.18
(therefore including upper limits on the radio flux) is 4.74 with
interquartile 0.12.  In the same redshift-luminosity bin there are 25
spectroscopically identified type 1 QSOs in the COSMOS bright MIPS
sample, all with X-ray direct detection and eight with a radio flux.
Their median radio luminosity (including limits on the radio
luminosity) and the median log ratio between the infrared and radio
luminosity (including limits on the radio luminosity) are 40.19 and
5.07 respectively. The probability that the luminosity ratio
distribution of the 38 sources in cells A and B and the 25 type 1 QSO
are drawn from the same distribution is 0.002\%, using the F test.
This would suggest a slightly stronger radio emission in infrared
selected CT QSOs than in unobscured, type 1 QSOs, qualitatively in
agreement with our predictions. However, we note that different
components, in addition to star-formation, can contribute to the
observed radio flux. Furthermore, some residual extinction may still
reduce the 8-24$\mu$m flux of the CT AGN, which would then
underestimate the true 5.8$\mu$m rest frame luminosity of these
sources. For all these reasons we consider the lower infrared to radio
luminosity ratio of CT AGN with respect to type 1 AGN certainly
intriguing but not conclusive.

A cleaner test of these predictions can come from infrared
spectroscopy of PAH features.  Fortunately, the COSMOS infrared
selected, CT AGN are bright enough to provide relatively high signal
to noise {\it Spitzer} IRS spectra.  These spectra will be able to
assess whether nuclear activity and strong star-formation are present
at the same time in these object, thus validating or disproving our
feedback scenario for AGN obscuration. Finally, if strong
star-formation is present in the host galaxies of the CT QSOs, as
expected, they should stand out in forthcoming deep Herschel surveys
at 70 and 110$\mu$m.

\section{Conclusions}

We found that 25--30\% of the MIPS 24$\mu$m sources brighter than
550$\mu$Jy have a direct X-ray detection down to an X-ray flux limit
of a few$\times10^{-16}$ \cgs. About 75\% of these sources are likely
to be AGN with L(2-10)keV$>10^{42}$ ergs s$^{-1}$.
We evaluated the fraction of obscured AGN in the COSMOS MIPS sample without
a direct X-ray detection by comparing the count rates and hardness ratio
in stacked X-ray images with detailed Monte Carlo simulations. 
We found that the fraction of AGN in this MIPS sample
(both X-ray detected and recovered through their infrared/optical
color and a detailed X-ray stacking analysis) is $49\pm10\%$.
Considering only the sources with z$>0.6$ the fraction increases to
$0.67\pm0.06$.  This is significantly higher than previous estimates
obtained using a much shallower X-ray coverage and an analysis of the
24$\mu$m-8$\mu$m color.

We computed the volume density of the MIPS 24$\mu$m selected sources
into two luminosity-redshift bins and corrected it for the fraction of
CT AGN found in nine cells defined in the F(24$\mu$m)/F(R) -- R-K
diagram, to find the volume density of infrared selected, CT
AGN. Our analysis shows that deep X-ray data are the key element
to obtain complete unbiased AGN samples, both through direct detection
and through dedicated stacking analyses. Of course the latter can
provide results on CT AGN valid only in a statistical sense. While the
search for and characterization of CT AGN remains one of the main
goals of the on-going {\it Chandra} and {\it XMM-Newton} ultra-deep
surveys of the CDFs (e.g. Comastri \& Brusa 2007), the direct X-ray
detection of large samples of CT AGN and the accurate measure of their
obscuring column densities must await for the next generation of X-ray
telescopes with imaging capabilities in the 10-100 keV band, like
NuSTAR, NeXT and Simbol-X (Fiore et al. 2008b).

We found that the density of CT QSOs with z=1.2--2.2 and log$\lambda
L_\lambda (5.8\mu$m)$=44.79-46.18$ and with z=0.7--1.2 and log$\lambda
L_\lambda (5.8\mu$m)$=44.06-44.79$ are $\sim 44\%$ and $\sim 67\%$ of
the density of X-ray selected unobscured and moderately obscured AGN
in the same redshift and luminosity bins, respectively.

Our results imply that the correlation between the fraction of
obscured AGN with the luminosity found for X-ray selected AGN holds
also when considering infrared selected, CT AGNs. If the fraction of
obscured AGN is a measure of the timescale over which the nuclear
feedback is at work, then unobscured and moderately obscured QSOs
should be hosted in more passive galaxies, on average, than those
hosting CT QSOs of similar luminosity. Star-formation can be traced at
radio wavelengths and we find indeed that the infrared selected CT
QSOs at z=1.2--2.2 are more radio luminous (with respect to their
5.8$\mu$m luminosity) than unobscured type 1 QSOs of similar redshift
and luminosity. Although this result is in line with previous findings
we do not consider it as conclusive.  Further investigation, as for
example direct infrared spectroscopy and far infrared photometry of
the candidate CT QSOs, is needed to confirm our evolutionary feedback
scenario.

\acknowledgments
We acknowledge support from ASI/INAF contracts
I/023/05/0 and I/024/05/0 and by PRIN/MUR grant 2006-02-5203. 
FF thanks Mari Polletta for useful comments. 
The zCOSMOS ESO Large Program Number 175.A-0839 is acknowledged.

\end{document}